\documentclass[12pt]{article}
\usepackage{graphicx}
\usepackage{amsmath}

\usepackage{amscd,amssymb}
\usepackage[cmtip,arrow,curve,matrix]{xy}

\newtheorem{Lemma}[equation]{Lemma}
\newtheorem{Proposition}[equation]{Proposition}

\newtheorem{Example}[equation]{Example}

\DeclareMathOperator{\ch}{ch}
\newcommand{\C}{\mathbb{C}}
\newcommand{\cE}{\mathcal{E}}
\newcommand{\cF}{\mathcal{F}}

\newcommand{\cG}{\mathcal{G}}

\newcommand{\heq}{\approx}

\newcommand{\iso}{\cong}
\DeclareMathOperator{\Img}{Img}

\DeclareMathOperator{\Ker}{Ker}
\DeclareMathOperator{\Coker}{Coker}

\renewcommand{\O}{\mathcal{O}}

\newcommand{\psb}[1]{[ \! [#1] \! ]}

\newcommand{\ptspace}{\ast}

\renewcommand{\P}{\mathbb{P}}
\newcommand{\Q}{\mathbb{Q}}

\newcommand{\R}{\mathbb{R}}
\newcommand{\redE}{\Bar{E}}
\newcommand{\restr}[1]{|_{#1}}

\newcommand{\Smash}{\wedge}

\DeclareMathOperator{\Td}{Todd}

\newcommand{\uln}[1]{\underline{#1}}

\newcommand{\xra}[1]{\xrightarrow{#1}}

\newcommand{\Z}{\mathbb{Z}}

\begin{document}

\hfill VPI-IPNAS-09-02

\vspace{1.0in}

\begin{center}
{\large\bf Elliptic genera of Landau-Ginzburg models over nontrivial spaces}

\vspace{0.5in}

Matt Ando$^1$, Eric Sharpe$^2$

\begin{tabular}{cc}
  \begin{tabular}{rl}
         $^1$ \hspace*{-5.5mm} & Department of Mathematics\\
                               & University of Illinois, Urbana-Champaign \\
                               & 1409 West Green Street \\
                               & Urbana, IL  61801\\
  \end{tabular} &
  \begin{tabular}{rl}
         $^2$ \hspace*{-5mm} & Physics Department\\
                             & Robeson Hall (0435) \\
                             & Virginia Tech\\
                             & Blacksburg, VA  24061\\
  \end{tabular}
\end{tabular}

{\tt mando@math.uiuc.edu}, {\tt ersharpe@vt.edu}

$\,$

\end{center}

In this paper, we discuss elliptic genera of (2,2) and (0,2)
supersymmetric Landau-Ginzburg models
over nontrivial spaces, {\it i.e.} nonlinear sigma models on
nontrivial noncompact manifolds with superpotential,
generalizing old computations in Landau-Ginzburg
models over (orbifolds of) vector spaces.  For Landau-Ginzburg models
in the same universality class as
nonlinear sigma models, we explicitly check 
that the elliptic genera of the Landau-Ginzburg
models match that of the nonlinear sigma models, via a Thom class computation
of a form analogous to that appearing in recent
studies of other properties
of Landau-Ginzburg models on nontrivial spaces.

\begin{flushleft}
May 2009
\end{flushleft}

\newpage

\tableofcontents

\newpage

\section{Introduction}

Historically, elliptic genera have provided an important example
of mathematics/physics interactions.  For mathematicians, elliptic genera
(and elliptic cohomology) provided the promise of new mathematical
invariants associated to spaces.  For physicists, elliptic genera are not
only one-loop string partition functions, but have also proven their
worth through {\it e.g.} their application to black hole entropy
computations \cite{sv}.

Elliptic genera of (2,2) supersymmetric Landau-Ginzburg models
over vector spaces were first computed in \cite{edlg}.
As many Landau-Ginzburg models are on the same K\"ahler moduli space
as ordinary nonlinear sigma models, and elliptic genera are invariant
under smooth deformations, computations of elliptic genera of Landau-Ginzburg
models often provide efficient ways to compute elliptic genera of
corresponding nonlinear sigma models.  Unfortunately, most of those
Landau-Ginzburg models do not live over vector spaces, or orbifolds
thereof, but over more complicated spaces, forming what are sometimes
called ``hybrid Landau-Ginzburg models.''

In this paper we shall generalize the methods of \cite{edlg} to 
Landau-Ginzburg models over nontrivial spaces, typically, total spaces
of vector bundles.  We study both (2,2) supersymmetric examples,
as well as more general (0,2) supersymmetric examples pertinent to
heterotic strings.  We check our methods by using the renormalization
group:  if a Landau-Ginzburg model is in the same universality class
as an ordinary nonlinear sigma model, then their elliptic genera must
match, and we verify this in our examples.  We will find that
in such cases, the two expressions for the elliptic genus are related
by a Thom class, a mathematical gadget that encodes how 
genera on one space can be
calculated in a larger space in which the
first space is embedded.

Much of this paper can also be seen as a step in a larger program
of generalizing computations for Landau-Ginzburg models on
vector spaces, to Landau-Ginzburg models on nontrivial spaces.
Other steps in this direction were in, for example,
\cite{alg22,alg02}, where it was described how to compute correlation
functions in A and B twisted Landau-Ginzburg models on nontrivial
spaces.  There, results were checked by comparing Landau-Ginzburg models
to nonlinear sigma models in the same universality class:
the resulting correlation functions were isomorphic, as expected,
and the computations in Landau-Ginzburg models gave a physical realization
of tricks for computing Gromov-Witten invariants, for example.
Other steps towards understanding Landau-Ginzburg models on nontrivial
spaces were described in \cite{clarke}, where mirror symmetry
was described as a duality between Landau-Ginzburg models on nontrivial
spaces, generalizing other approaches to the subject.

We begin in section~\ref{review} with a review of elliptic genus computations
for nonlinear sigma models and Landau-Ginzburg models over vector spaces.
In section~\ref{nontriv-spaces} we generalize both of those computations,
to compute elliptic genera of both (2,2) and (0,2) supersymmetric
Landau-Ginzburg models over nontrivial spaces.  
As a consistency check, we compare elliptic genera of Landau-Ginzburg
models to elliptic genera of nonlinear sigma models in the same
universality class.  As elliptic genera are indices, they are invariant
under renormalization group flow, and using results proven in 
appendix~\ref{app:thom}, we explicitly verify that elliptic genera of
theories in the same universality class do match.
Mathematically, that matching is realized via ``Thom classes,'' which
also appeared, in another form, in \cite{alg22,alg02}.
In section~\ref{thom-genl} we conclude with a general discussion of
Thom classes and their appearance in physics.
In appendix~\ref{app:identities} we list some handy identities for
manipulating elliptic genera, and in appendix~\ref{app:thom} we derive
identities for Thom classes in elliptic genera that are used in the
bulk of the text. 

As the consistency checks of this paper rely on one example of
a Landau-Ginzburg/Calabi-Yau correspondence (albeit not the usual
one), let us mention for completeness the works \cite{ej,gm} which
consider elliptic genera in the context of different Landau-Ginzburg
models and a different Landau-Ginzburg/Calabi-Yau correspondence.

\section{Review}   \label{review}

\subsection{Elliptic genera of nonlinear sigma models}
\label{review-nlsm}

Elliptic genera of nonlinear sigma models have been discussed extensively
elsewhere, so we shall review them only briefly.

An elliptic genus is, physically, the one-loop partition function of a 
theory with at least (0,2) supersymmetry
in which the right-moving fermions are all in a R sector --
equivalently, the partition function of a half-twisted theory -- and,
possibly, the left-moving states are also twisted in some way.
More specifically, we shall consider elliptic genera which are of the form
\begin{equation}    \label{eg-nlsm-def1}
\mbox{Tr } (-)^{F_R} \exp(i \gamma J_L) q^{L_0} \overline{q}^{
\overline{L}_0}
\end{equation}
where $q$ is the modular parameter, and the current $J_L$ is a left-moving
$U(1)$ current which is implicitly assumed to exist.

Computation of such genera has been discussed in many places in the
literature, beginning in \cite{edoldeg1}, but let us take a few moments
to review the highlights.
We shall consider theories of the
form of  
nonlinear sigma models with (0,2) supersymmetry, defined on a complex
K\"ahler manifold $X$ of dimension $n$ with a gauge bundle ${\cal E}$
of rank $r$ satisfying
\begin{equation}\label{eq:33}
\begin{array}{c}
\Lambda^{top} {\cal E} \: = \: K_X \\
{\rm ch}_2(TX) \: = \: {\rm ch}_2({\cal E})
\end{array}
\end{equation}
In addition, we shall usually assume $X$ is Calabi-Yau (though we
shall note special cases in which sensible results can be obtained
more generally).
It can be shown \cite{edoldeg1,edoldeg2}
that equation~(\ref{eg-nlsm-def1}) is an index,
and so is invariant under smooth deformations of the theory.
As a result, we can consistently deform the theory to the large-radius
limit, where the computation of~(\ref{eg-nlsm-def1}) becomes a free-field
computation.

Because the right-movers are all in the R sector, the nonzero modes of the
right-moving fermions and bosons cancel out, leaving only the left-movers
and right-moving zero modes to contribute.  The right-moving zero modes
are defined by a Fock vacuum transforming as a spinor lift\footnote{
Spinor lifts do not always exist; however, sometimes it is still possible
to make sense of such expressions.  We shall speak more about this as various
special cases arise. 
} of $TX$.  As a result, all of the states appearing in the
trace~(\ref{eg-nlsm-def1}) have spinor indices.  In particular,
the trace~(\ref{eg-nlsm-def1}) is the index of the Dirac operator
coupled to various bundles defined by the nonzero modes of the fields.

To make this a little more concrete,
below we list bosonic oscillators at a few mass levels and 
corresponding bundles, for a nonlinear sigma model on $X$:
\begin{center}
\begin{tabular}{ccc}
Mass level & Oscillator & Bundle \\ \hline
1 & $\alpha^{\mu}_{-1}$ & $TX$ \\
2 & $\alpha^{\mu}_{-2}$, $\alpha^{\mu}_{-1} \alpha^{\mu}_{-1}$ &
$TX \oplus \mbox{Sym}^2 (TX)$ \\
3 & $\alpha^{\mu}_{-3}$, $\alpha^{\mu}_{-2} \alpha^{\nu}_{-1}$,
$\alpha^{\mu}_{-1} \alpha^{\nu}_{-1} \alpha^{\rho}_{-1}$ &
$TX \oplus \left( TX \otimes TX \right) \oplus
\mbox{Sym}^3(TX)$
\end{tabular}
\end{center}
At mass level $n$, it is straightforward to check that the bundle
obtained above is the coefficient of $q^n$ in the following
element of the Grothendieck group of vector bundles:
\begin{equation}   \label{sym-tensor1}
\bigotimes_{n=1,2,3,\cdots} S_{q^n}(TX)
\end{equation}
where
\begin{displaymath}
S_q(TX) \: = \: 1 \: + \: q TX \: + \:
q^2 \mbox{Sym}^2 (TX) \: + \:
q^3 \mbox{Sym}^3 (TX) \: + \: \cdots
\end{displaymath}
Each factor of $S_{q^n}(TX)$ corresponds to a set of states
of the form
\begin{displaymath}
\left\{1, \alpha^{\mu}_{-n}, \alpha^{\mu_1}_{-n} \alpha^{\mu_2}_{-n},
\alpha^{\mu_1}_{-n} \alpha^{\mu_2}_{-n} \alpha^{\mu_3}_{-n}, \cdots
\right\}
\end{displaymath} 
and so the tensor product encodes all products of all nonzero
oscillator creation operators.
Thus, for example, the final result for the elliptic genus will involve
computing the index of a bundle which has, among other things, a factor
of the tensor product~(\ref{sym-tensor1}).
Furthermore, because we have been implicitly working with complex
manifolds, holomorphic bundles, and we distinguish $\alpha_{-1}^i$
from $\alpha_{-1}^{\overline{i}}$, we have
\begin{equation}   \label{sym-tensor2}
\bigotimes_{n=1,2,3,\cdots} S_{q^n}\left((TX)^{\bf C} \equiv
TX \oplus \overline{TX}\right)
\end{equation}

In the (0,2) supersymmetric nonlinear sigma models
we consider, the current $J_L$
exists by virtue of the condition
$\Lambda^{top} {\cal E} \cong {\cal O}_X$ on ${\cal E}$
(and becomes the left R-current in the special case of (2,2) supersymmetry).
If the left-moving fermions are in an NS sector, then 
the trace~(\ref{eg-nlsm-def1}) is given by \cite{edoldeg1,edoldeg2}
\begin{equation}
q^{-(1/24)(2n + r)}
\int_X \Td(TX) %\hat{A}(TX) 
\wedge {\rm ch}\left( \bigotimes_{n=1,2,3,\cdots}
S_{q^n}((TX)^{\bf C}) 
\bigotimes_{n=1/2,3/2,5/2,\cdots}
 \Lambda_{q^n}\left((e^{i \gamma} {\cal E})^{\bf C}\right)
\right)
\end{equation}
(compare {\it e.g.} \cite{edoldeg1}[equ'n (30)])
where $S_q(TX)$ is as above, 
$\Lambda_q({\cal E})$ denotes an element of the Grothendieck
group of vector bundles on $X$ defined analogously as the linear combinations
\begin{eqnarray*}
%S_q(TX) & = &
%1 \: + \: q TX \: + \: q^2 \mbox{Sym}^2 (TX) \: + \: q^3
%\mbox{Sym}^3(TX) \: + \:
%\cdots \\
\Lambda_q({\cal E}) & = &
1 \: + \: q {\cal E} \: + \: q^2 \mbox{Alt}^2({\cal E}) \: + \:
q^3 \mbox{Alt}^3({\cal E}) \: + \: \cdots
\end{eqnarray*}
(arising physically from the left-moving fermion oscillator modes,
just as the factor~(\ref{sym-tensor1}) arose from bosonic oscillator
modes),
and the ${\bf C}$ symbol indicates complexification:
\begin{eqnarray*}
(TX)^{\bf C} & = & T^{1,0}X \oplus \overline{ T^{1,0}X}
\\
(z{\cal E})^{\bf C} & = & z{\cal E} \oplus \overline{z} \overline{{\cal E}}
\end{eqnarray*}
The prefactor of $q$ is due to the zero energy of the vacuum:
each periodic complex boson contributes $-1/12$, and each
antiperiodic complex fermion contributes $-1/24$. 
The fact that the $S_{q^n}$'s are tensored together for integer $n$
reflects the fact that the bosonic oscillators are integrally moded;
the fact that the $\Lambda_{q^n}$'s are tensored together for
half-integer $n$'s reflects the fact that the fermionic oscillators
are half-integrally moded.

If the left-moving fermions are in a R sector rather than a NS sector,
then the elliptic genus 
\begin{displaymath}
\mbox{Tr}_{{\rm R},{\rm R}} (-)^{F_R} \exp\left( i \gamma J_L \right)
q^{L_0} \overline{q}^{\overline{L}_0}
\end{displaymath}
is given by
\begin{eqnarray} \label{eq:32}
\lefteqn{
q^{+(1/12)(r - n)}
} \nonumber \\
& & \cdot \int_X \hat{A} (TX) % \Td (TX)   %\hat{A}(TX) 
\wedge {\rm ch}\Biggl( 
%\left( {\cal S}_+({\cal E}^{\vee}) \oplus {\cal S}_-({\cal E}^{\vee}) 
%\right)
z^{-r/2} \left( \det {\cal E} \right)^{+1/2}
\Lambda_1\left( z {\cal E}^{\vee} \right)
\nonumber \\
& & \hspace*{1.5in} \left. \cdot \bigotimes_{n=1,2,3,\cdots}
S_{q^n}((TX)^{\bf C}) 
\bigotimes_{n=1,2,3,\cdots}
 \Lambda_{q^n}\left((z^{-1} {\cal E})^{\bf C}\right)
\right)
\end{eqnarray}
where $z = \exp(-i \gamma)$.
(Compare {\it e.g.} \cite{edoldeg2}[equ'n (31)].)
In the case $X$ is Calabi-Yau, $\det {\cal E}$ is trivial, since
$\Lambda^{\rm top} {\cal E} \cong K_X$, so the expression above is
well-defined.

The factor of 
\begin{equation}   \label{fock-vacua-reln}
z^{-r/2} \left( \det {\cal E} \right)^{+1/2} \Lambda_1( z {\cal E}^{\vee} )
\: = \:
z^{+r/2} \left( \det {\cal E} \right)^{-1/2} \Lambda_1( z^{-1} {\cal E} )
\end{equation}
arises above from the zero modes of the left-moving fermions.
It reflects the ambiguity in the Fock vacuum:
if we define $|0\rangle$ by $\lambda_-^a |0\rangle = 0$,
then we have a set of vacua
\begin{displaymath}   
| 0 \rangle, \:
\lambda_-^{\overline{a}} | 0 \rangle, \:
\cdots, 
\lambda_-^{\overline{a}_1} \cdots \lambda_-^{\overline{a}_r}
| 0 \rangle
\end{displaymath}
Similarly, if instead we define $| 0 \rangle$ by
$\lambda_-^{\overline{a}} | 0 \rangle = 0$,
then we have an equivalent set of vacua
\begin{displaymath}
| 0 \rangle, \:
\lambda_-^{a} | 0 \rangle, \:
\cdots, 
\lambda_-^{a_1} \cdots \lambda_-^{a_r}
| 0 \rangle
\end{displaymath}
The existence of these two equivalent characterizations of the Fock
vacua corresponds to the two sides of equation~(\ref{fock-vacua-reln}).
Furthermore, these vacua correspond to spinor lifts of ${\cal E}$: 
note that we can write
\begin{displaymath}
\Lambda_1( z {\cal E}^{\vee} ) \: = \:
{\cal S}_+(z {\cal E}^{\vee}) \oplus {\cal S}_-(z{\cal E}^{\vee}) 
\end{displaymath}
where 
${\cal S}_{\pm}$ denote the two chiral Spin$^c$ lifts
of ${\cal E}^{\vee}$, {\it i.e.}
\begin{eqnarray*}
{\cal S}_+({\cal E}^{\vee}) & \equiv & \bigoplus_{n \: {\rm even}} 
\Lambda^n {\cal E}^{\vee}\\
{\cal S}_-({\cal E}^{\vee}) & \equiv & \bigoplus_{n \: {\rm odd}}
\Lambda^n {\cal E}^{\vee} 
\end{eqnarray*}
which are made into honest spinors via the $\sqrt{ \det {\cal E} }$
factors.
(Physically, every vector bundle comes with a hermitian fiber metric,
so we will often fail to distinguish ${\cal E}^{\vee}$ from
$\overline{ {\cal E} }$.)
The prefactor of $q$
is due to the vacuum zero energy:  each periodic complex boson contributes
$-1/12$, and each periodic complex fermion contributes $+1/12$.

Note that in the spinor lifts of ${\cal E}$, ${\cal E}$ is not 
complexified, unlike the nonzero modes.  This is because for the
ambiguity in the Fock vacuum, we use the relation $\{ \psi_0^i,
\psi_{0 j} \} \propto \delta^i_j$, so we take one of either
$\psi_0^i$, $\psi_{0 j}$ to be creation operators and the other to
be annihilation operators -- the choice does not matter, as the resulting
collection of states are the same.

Readers familiar with elliptic genera computations elsewhere
should note that the ``Witten genus'' can be obtained as a special
case of the R sector genus above.  Specifically, for $z=-1$,
the R sector genus above is proportional to
\begin{eqnarray} \label{eq:31}
\lefteqn{
{\rm Tr}_{R,R} (-)^{F_R} (-)^{F_L} 
q^{L_0} \overline{q}^{\overline{L}_0}
} \nonumber \\
&=  & q^{+(1/12)(r - n)} \int_X \hat{A} (TX) % \Td (TX)   %\hat{A}(TX) 
\wedge {\rm ch}\Biggl( 
 \left( \det {\cal E} \right)^{+1/2}
\Lambda_{-1}\left(  {\cal E}^{\vee} \right)
\nonumber \\
& & \hspace*{1.75in} \left. \cdot \bigotimes_{n=1,2,3,\cdots}
S_{q^n}((TX)^{\bf C}) 
\bigotimes_{n=1,2,3,\cdots}
 \Lambda_{-q^n}\left(( {\cal E})^{\bf C}\right)
\right)
\end{eqnarray}
which is precisely the Witten genus, introduced
by Witten in \cite{edoldeg1}[equ'n (31)].
This genus has been shown to play a
fundamental role in elliptic cohomology \cite{AHS01}.  When
$z=1$, both the genus \eqref{eq:32} and \eqref{eq:31} are modular
provided the two conditions   \eqref{eq:33} hold; it is not\footnote{
Physically, for non-Calabi-Yau cases, one can still imagine computing
the genus {\it at}, though not away from,
the extreme large-radius free-field limit of the
nonlinear sigma model.  The left-moving $U(1)$ current will no longer
be nonanomalous, but if we fix $z = \pm 1$, then that is not a concern.
} necessary 
to require that $X$ be Calabi-Yau.

\subsection{Elliptic genera of Landau-Ginzburg models over
vector spaces}

\subsubsection{Physical analysis -- R sector}

Later in this paper we shall compute
elliptic genera of Landau-Ginzburg models over topologically
nontrivial spaces.  We have already reviewed elliptic genus computations
in nonlinear sigma models; next, let us review the computation of
elliptic genera in Landau-Ginzburg models on topologically trivial
spaces, with quasi-homogeneous superpotentials,
as first discussed in \cite{edlg}.  In particular, we will focus
on the special case of a Landau-Ginzburg model over the complex
line, with a monomial superpotential.

Recall that in that paper, a Landau-Ginzburg model over the complex
line ${\bf C}$ was considered, with superpotential $W = \Phi^{k+2}$.
The elliptic genus was defined there as the trace
\begin{displaymath}
\mbox{Tr } (-)^{F_R} q^{L_0} \overline{q}^{\overline{L}_0}
\exp(i \gamma J_L)
\end{displaymath}
over states in which all fields have R boundary conditions 
along spacelike directions:
\begin{eqnarray*}
\phi(x_1+1,x_2) & = & \phi(x_1,x_2) \\
\psi_+(x_1+1,x_2) & = &  \psi_+(x_1,x_2) \\
\psi_-(x_1+1,x_2) & = &  \psi_-(x_1,x_2) 
\end{eqnarray*}
The boundary conditions along timelike directions require more explanation.
First, let us work out the left R-charges of the fields, so as to 
understand the $\exp(i \gamma J_L)$ factor in the trace.
Because of the superpotential interactions, the left R-symmetry no longer
merely rotates the $\psi_-$'s by a phase, leaving other fields invariant,
but rather rotates all of the fields by some phase.
It is straightforward to check that the
left R-charges are as follows:
\begin{center}
\begin{tabular}{cc}
Field & R-charge \\ \hline
$\phi$ & $1$ \\
$\psi_+$ & $1$ \\
$\psi_-$ & $-(k+1)$ \\
\end{tabular}
\end{center}

Furthermore, also because of the superpotential interactions,
$(-)^{F_R}$ no longer merely corresponds to a sign on $\psi_+$'s;
rather, it generates a sign on both $\psi_+$ and $\psi_-$ simultaneously.

We do not list here the timelike boundary conditions, but the 
attentive reader should recall, for example, that
fields with Ramond boundary conditions along
timelike directions correspond to traces with $(-)^F$ factors.

The zero modes of $\psi_-$ contribute a factor of
\begin{displaymath}
\exp(- i \gamma(k+1)/2 ) \: - \: \exp( +i \gamma (k+1)/2 )
\end{displaymath}
the zero modes of $\psi_+$ contribute a factor of
\begin{displaymath}
\exp(i \gamma/2) \: - \: \exp(- i \gamma/2)
\end{displaymath}
The nonzero modes of the fermions contribute
\begin{displaymath}
\prod_{n=1}^{\infty}
%\left( 1 \: - \: q^n \exp(-i \gamma (k+1) ) \right)
\left( 1 \: - \: z^{k+1} q^n  \right)
%\left( 1 \: - \: q^n \exp(i \gamma(k+1) ) \right)
\left( 1 \: - \: z^{-(k+1)} q^n  \right)
%\left( 1 \: - \: \overline{q}^n \exp(i \gamma) \right)
\left( 1 \: - \: z^{-1} \overline{q}^n  \right)
%\left( 1 \: - \: \overline{q}^n \exp(- i \gamma) \right)
\left( 1 \: - \: z \overline{q}^n  \right)
\end{displaymath}
where
\begin{displaymath}
z \: = \: \exp(-i \gamma )
\end{displaymath}
and
where the minus signs are due to the $(-)^{F_R}$ factor in the trace
(and the fact that because of the superpotential interactions,
$(-)^{F_R}$ multiplies both $\psi_-$ and $\psi_+$ simultaneously by
a sign).
We can rewrite this in the form of the index of a Dirac operator.
Using the notation of appendix~\ref{app:identities}, if we let $L$ denote
the tangent bundle of ${\bf C}$ restricted to the origin, then the expression
above for the contribution from the nonzero modes of the fermions is of
the form
\begin{displaymath}
\mbox{ch}\left(
\bigotimes_{n=1,2,3,\cdots} \Lambda_{-q^n}\left((z^{k+1} L)^{\bf C}
\right) 
\bigotimes_{n=1,2,3,\cdots} \Lambda_{-\overline{q}^n} \left(
 ( z^{-1} L )^{ {\bf C} } \right)
\right)
\end{displaymath}

The nonzero modes of the bosons contribute
\begin{displaymath}
\prod_{n=1}^{\infty}
\frac{1}{1 \: - \: z^{-1} q^n  }
\frac{1}{1 \: - \: z q^n  }
\frac{1}{1 \: - \: z^{-1} \overline{q}^n  }
\frac{1}{1 \: - \: z \overline{q}^n  }
\end{displaymath}
which we can rewrite as
\begin{displaymath}
\mbox{ch}\left(
\bigotimes_{n=1,2,3,\cdots} S_{q^n}\left( ( z^{-1} L )^{ {\bf C} } 
\right)
\bigotimes_{n=1,2,3,\cdots} S_{\overline{q}^n}\left(
( z^{-1} L )^{\bf C} \right)
\right)
\end{displaymath}
Note that the $\overline{q}$ contributions from the bosons and fermions
cancel out -- this can be seen either directly from the product formula above
or using the identities in appendix~\ref{app:identities}.

Finally, in this special case,
the zero modes\footnote{
Taking into account the timelike boundary conditions, there are no
zero modes from the point of view of path integral quantization.
The $\phi_0$, $\overline{\phi}_0$ referred to here are solely an artifact
of the periodic moding in canonical quantization.
}
 $\phi_0$, $\overline{\phi}_0$ also contribute
a factor of
\begin{displaymath}
\left(
\frac{1}{1 \: - \: z} 
\right) \left(
\frac{1}{1 \: - \: z^{-1} }
\right)
\end{displaymath}
as discussed in \cite{edlg}.

Putting this together, we get the genus
\begin{equation}\label{eq:9}
f (z,q) = \frac{ z^{-1/2} \left( z^{(k+1)/2}  \: - \: z^{-(k+1)/2} 
\right)
%\left( z^{-1/2} \: - \: z^{+1/2}
%\right)
}{
\left( 1 \: - \: z^{-1} \right)
}
\prod_{n=1}^{\infty} \frac{
\left( 1 \: - \: z^{k+1} q^n   \right)
\left( 1 \: - \: z^{-(k+1)} q^n  \right)
}{
\left( 1 \: - \: z^{-1} q^n  \right)
\left( 1 \: - \: z q^n \right)
}
\end{equation}

We can interpret this as index theory over a point, the fixed-point locus
of a $U(1)$ action, {\it i.e.} the space
of bosonic zero modes satisfying the boundary condition
\begin{displaymath}
\phi(x_1,x_2+1) \: = \: \exp(i \gamma) \phi(x_1,x_2)
\end{displaymath}
for generic $\gamma$,
which is to say, $\phi_0 = \{ 0 \}$.

Note that the expression for the nonzero
modes is given by
\begin{displaymath}
\mbox{ch} \left(
\bigotimes_{n \geq 1} S_{z^{-1} q^n }(L) \otimes S_{z q^n }(
\overline{L} )
\otimes \Lambda_{- z^{k+1} q^n }(L)
\otimes \Lambda_{- z^{-(k+1)} q^n  }(\overline{L})
\right)
\end{displaymath}

Instead of working with fields with R boundary conditions along spacelike
directions, one could instead try to 
compute elliptic genera in which the $\psi_-$
have NS boundary conditions in spacelike directions.  Note, however,
that because of the $\psi_+ \psi_- \phi^k$ Yukawa coupling in the theory,
if the $\psi_-$ have NS boundary conditions, then so too must the $\psi_+$,
and then the right-moving contributions would no longer cancel out.

\subsubsection{Physical analysis -- NS sector}

In the last subsection we reviewed the results of \cite{edlg} on computing
elliptic genera of Landau-Ginzburg models over vector spaces,
in the R sector.  In this subsection we will extend the results of
\cite{edlg} to the NS sector.

From the table of left R-charges in the last subsection, 
we see that fields in the NS sector have spacelike boundary conditions
\begin{eqnarray*}
\phi(x_1+1,x_2) & = & - \phi(x_1,x_2) \\
\psi_+(x_1+1,x_2) & = & - \psi_+(x_1,x_2) \\
\psi_-(x_1+1,x_2) & = & (-)^{k+1} \psi_-(x_1,x_2)
\end{eqnarray*}

From the spacelike boundary conditions above, we see that we must
consider the cases of $k$ even and odd separately.

In the case $k$ is even, there are no zero modes at all.
The fermions contribute
\begin{displaymath}
\prod_{n=1/2,3/2,\cdots} \left[ 1 - z^{k+1} q^n \right]
\left[ 1 - z^{-(k+1)} q^n \right]
\left[ 1 - z^{-1} \overline{q}^n \right]
\left[ 1 - z \overline{q}^n \right]
\end{displaymath}
and the bosons contribute
\begin{displaymath}
\prod_{n=1/2,3/2,\cdots}
\left[ 1 - z^{-1} q^n \right]^{-1}
\left[ 1 - z q^n \right]^{-1}
\left[ 1 - z^{-1} \overline{q}^n \right]^{-1}
\left[ 1 - z \overline{q}^n \right]^{-1}
\end{displaymath}
Putting this together, we see that for $k$ even, the elliptic genus is
given by
\begin{displaymath}
\prod_{n=1/2,3/2,\cdots}
\left[ 1 - z^{k+1} q^n \right]
\left[ 1 - z^{-(k+1)} q^n \right]
\left[ 1 - z^{-1} q^n \right]^{-1}
\left[ 1 - z q^n \right]^{-1}
\end{displaymath}

In the case $k$ is odd, there are $\psi_-$ zero modes.
In this case, the total contribution from the fermions is
\begin{displaymath}
\left( z^{(k+1)/2} - z^{-(k+1)/2} \right)
\prod_{n=1,2,\cdots}
\left[ 1 - z^{k+1} q^n \right]
\left[ 1 - z^{-(k+1)} q^n \right]
\prod_{n=1/2,3/2,\cdots}
\left[ 1 - z^{-1} \overline{q}^n \right]
\left[ 1 - z \overline{q}^n \right]
\end{displaymath}
and the bosons contribute
\begin{displaymath}
\prod_{n=1/2,3/2,\cdots}
\left[ 1 - z^{-1} q^n \right]^{-1}
\left[ 1 - z q^n \right]^{-1}
\left[ 1 - z^{-1} \overline{q}^n \right]^{-1}
\left[ 1 - z \overline{q}^n \right]^{-1}
\end{displaymath}
Putting this together, we see that for $k$ odd, the elliptic genus is
given by
\begin{displaymath}
\left( z^{(k+1)/2} - z^{-(k+1)/2} \right)
\prod_{n=1,2,\cdots}
\left[ 1 - z^{k+1} q^n \right]
\left[ 1 - z^{-(k+1)} q^n \right]
\prod_{n=1/2,3/2,\cdots}
\left[ 1 - z^{-1} q^n \right]^{-1}
\left[ 1 - z q^n \right]^{-1}
\end{displaymath}

\subsubsection{Mathematical interpretation}

It is striking that the Landau-Ginzburg elliptic genus \eqref{eq:9}
has a natural meaning in the equivariant elliptic cohomology of 
\cite{Gro07,Gre05}: it 
is the equivariant genus  or Euler class of a virtual representation of $U
(1),$ associated to the sigma orientation \cite{AHS01,And03,AG}.

Let $h$ be a generalized cohomology
theory.  By the suspension isomorphism, the reduced $h$-theory of the
$k$-sphere contributes to the cohomology of a point: precisely, we have
\[
    \tilde{h}^{0} (S^{k}) \iso h^{-k} (*).
\]
In stable homotopy theory, this statement even makes sense when $k$ is
negative.  

This has the following generalization in the equivariant setting.  
Let $G$ be a (compact Lie) group, and let $E$ be a $G$-equivariant
generalized cohomology theory.  If $V$ is a representation of $G$,
let $S^{V}$ be its one-point compactification.  Then 
\[
    \tilde{E}^{0} (S^{V}) \iso E^{-V} (*).
\]
Moreover, this statement makes sense even when $V$ is a virtual
representation of $G.$

Let $E$ be the $U (1)$-equivariant elliptic cohomology
associated to an elliptic curve $C$.  If $V$ is a representation of $U
(1)$, it turns out that $\tilde{E}^{0} (S^{V})\iso E^{-V} (*)$ is the
sections of a line bundle $\mathbb{L} (V)$ over the ellitpic curve
$C$.  More precisely, let $L_{k}$ be the 
one-dimensional representation of $U (1)$ given by $z\mapsto z^{k}$.
Let $C[k]$ be the divisor of points of order $k$ of $C$, and let 
$\O (-C[k]) = I (C[k])$ be the ideal sheaf of functions which vanish
to first order at $C[k]$.  Then 
one has \cite{Gre05}
\[
       \mathbb{L} (L_{k}) \iso \O (-C[k]),
\]
and 
\begin{equation}\label{eq:27}
      \mathbb{L} (L_{k+1}-L_{1}) \iso \O (-C[k+1]+C[1]) \iso \O
      (-[C[k+1]+ (0)).
\end{equation}
For convenience, let us write $V_{k}$ for the virtual representation
$L_{k}-L$, and 
let $D$ denote the divisor $(0)-C[k+1]$, so \eqref{eq:27} says
$\mathbb{L} (V_{k}) \iso \O (D)$.

As we recalled at \eqref{eq:31}, in \cite{edoldeg1}, Witten introduced
the genus of spin manifolds  
\[
   M\mapsto  \int \hat{A} (M) ch\left(\bigotimes_{n\geq 1} S_{q^{n}}
   (T^{\C}) \right);
\]
its $K$-theory characteristic series is
\begin{equation}\label{eq:30}
    \sigma (L,q) = (L^{1/2}-L^{-1/2}) \prod_{n\geq
    1}\frac{(1-q^{n}L) (1-q^{n}L^{-1})}{(1-q^{n})^{2}};
\end{equation}
In \cite{AHS01,AHR,Hop95,Hop02}, it was shown
that this genus plays a fundamental role in elliptic cohomology.
The expression \eqref{eq:30} gives an equivariant genus simply by
taking $L$ to be an 
equivariant line bundle, and in  \cite{And03,AG} it is shown that
this equivariant genus plays an equally fundamental role in
equviariant elliptic cohomology.  

Let $\tau$ be complex number with positive imaginary part, let
$\Lambda$ be the lattice $2\pi i \Z + 2 \pi i \tau \Z$, let $q=e^{2\pi i
\tau}$, and let $C$ be 
the elliptic curve 
\[
C=\C/\Lambda\iso \C^{\times}/q^{\Z}.
\]

Letting $z\in \C^{\times},$ the expression  $f (z,q)= \sigma
(z^{k},q)/\sigma (z,q)$ of \eqref{eq:9} is easily seen to be the
equivariant elliptic cohomology Euler class of the representation of
the virtual representation $L_{k+1}-L_{1}$ of $U(1),$ associated to
the characteristic series \eqref{eq:30}.  It
vanishes to first order at the points of order $k+1$ except the 
origin, so it is trying to be a section of $\O (D)$.  

The only apparent problem is that $f (z,q)$ does not descend to a
function on $C$.  Instead  
\begin{equation}\label{eq:28}
 f (zq^{n},q) = (-1)^{nk}
 z^{-n[(k+1)^{2}-1]}q^{-\tfrac{n^{2}}{2}[(k+1)^{2}-1]} 
 f (q,z).
\end{equation}
That is, $f$ is a section of the line bundle over $C$ given by this
transformation rule.  

This means that $f$ lives in the \emph{twisted}
equivariant elliptic cohomology of a point, and it is an important 
fact the twist is controlled by the second Stiefel
Whitney class and the first Pontrjagin class.  

Recall that if $u\in H^{2}(BU(1))$ is the generator, then a virtual
representation of $U(1)$ has equivariant Stiefel-Whitney, Chern, and
Pontrjagin classes which are elements of $H^{*} (BU(1))$ with $\Z/2$,
$\Z$, and $\Z$ coefficents.  One checks easily (recall that
$V_{k}=L_{k+1}-L_{1}$) that  
\begin{align*}
     w_{2}^{U(1)} (V_{k}) & \equiv k u \mod 2  \\
     p_{1}^{U(1)} (V_{k}) & = ((k+1)^{2}-1) u^{2}  = -\deg
     (D) u^{2}
\end{align*}

Thus we can rewrite the transformation rule \eqref{eq:28} as
\begin{equation}\label{eq:29}
     f (zq^{n},q) = (-1)^{nw_{2}^{U(1)} (V_{k})} z^{-np_{1}^{U(1)} (V_{k})}
     q^{-\tfrac{n^{2}}{2}p_{1}^{U(1)} (V_{k})} f (z,q).
\end{equation}

This is a general phenomenon: if $V$ is a virtual spin
$U(1)$-equivariant bundle over an $U(1)$-space $X$, then 
it is essentially a result of \cite{And03,AG} that we can use 
$\lambda (V)= \tfrac{p_{1}}{2} (V)$ to form a twisted form
of the equivariant elliptic cohomology of $X$, say $E
(X)_{\lambda (V)}$.  We can also form the reduced equivariant elliptic
cohomology of $X^{V}$, the one-point compactification of $V$, and
\[
     \tilde{E} (X^{V}) \iso E (*)_{\lambda (V)}.
\]
In the case $X$ is a point (and assuming $k$ even so we can ignore
$w_{2}$), this reduces to  
\[
     \tilde{E} (S^{V}) = E^{-V} (*) \iso E^{0}
     (*)_{\lambda (V)}. 
\]
Physically, the main observation is that the transformation rule
\eqref{eq:29} satisfied by the elliptic genus of the Landaug-Ginzburg
model is controlled by the equivariant $w_{2}$ and $p_{1}$ of the
representation $L_{k+1}-L_{1}$.

In the next section we shall consider more complicated examples of 
Landau-Ginzburg models.

\section{Landau-Ginzburg models over nontrivial spaces}
\label{nontriv-spaces}

In this section, we generalize both of the computations of the
previous section to discuss elliptic genera of Landau-Ginzburg
models over nontrivial spaces.  To compute elliptic genera of
Landau-Ginzburg models over nontrivial spaces, we need to assume
the spaces have ${\bf C}^{\times}$ actions with respect to which
the superpotential is quasi-homogeneous, much as in \cite{alg22,alg02}.
That ${\bf C}^{\times}$ action weights the contributions of the various
oscillator modes to the genus, via the $\exp(i \gamma J_L)$ factor
in the trace.  Moreover, because of the twisted boundary conditions,
the integral appearing in the genus, the integral over the space of
bosonic zero modes, will be an integral over the fixed-point locus of
that ${\bf C}^{\times}$ action.

In particular, we will primarily focus on total spaces of complex vector
bundles, for which the ${\bf C}^{\times}$ action in question will
be a rotation of the fibers.

We will check our results in cases that the Landau-Ginzburg models
are in the same universality classes as nonlinear sigma models,
by comparing elliptic genera.  The expressions we will derive for
genera in the two representatives of the universality class will typically
look different, but will turn out to match, often via Thom classes
(much as in A-twisted Landau-Ginzburg models \cite{alg22,alg02}).

\subsection{The (2,2) quintic and other complete intersections}    
\label{22quintic}

Consider a Landau-Ginzburg model over the space
\begin{displaymath}
X \: = \: \mbox{Tot} \left( {\cal O}(-5) \:
\stackrel{\pi}{\longrightarrow} \: {\bf P}^4 \right)
\end{displaymath}
with superpotential $W = p G$.
Using the same prescription as in \cite{edlg}, we would like to
compute the elliptic genus of this Landau-Ginzburg model.
Furthermore, since this Landau-Ginzburg model flows under the
renormalization group \cite{alg22} to a nonlinear sigma model on the quintic
$Q = \{ G = 0 \} \subset {\bf P}^4$, the elliptic genus of the
Landau-Ginzburg model should match that of the quintic.

First, let us review the elliptic genera of the quintic;
then, we shall compute the corresponding elliptic genera in the 
Landau-Ginzburg model, and check that they are the same.

Specializing the results in section~\ref{review-nlsm}, the elliptic genus
of the quintic, 
\begin{displaymath}
\mbox{Tr}_{{\rm NS},{\rm R}} 
(-)^{F_R} \exp(i \gamma J_L) q^{L_0} \overline{q}^{\overline{L}_0}
\end{displaymath}
with NS boundary
conditions along spacelike directions on the left-moving fermions, is
given by 
\begin{displaymath}
q^{-(1/24)(9)}
\int_Q \hat{A}(TQ) \wedge \mbox{ch}\left(
\bigotimes_{n=1,2,3,\cdots} S_{q^n} \left( (TQ)^{ {\bf C} } \right)
\bigotimes_{k=1/2,3/2,5/2,\cdots}
\Lambda_{q^k } \left( \left(e^{i \gamma} TQ\right)^{ {\bf C} } \right)
\right)
\end{displaymath}

Now, let us compute the corresponding elliptic genus in the
Landau-Ginzburg model, and check that they match.
In the case of the Landau-Ginzburg model, the left-moving $R$-charge
acts differently on the $p$ multiplet than on the rest of the fields.
From \cite{alg22}, the fields have left R-charges as follows:
\begin{center}
  \begin{tabular}{c|c|c|c}
         Field & $Q_L$  & Field & $Q_L$ \\ \hline
         $\phi^i$ & $0$ & $p$ & $-1$ \\
         $\psi_+^i$ & $0$ & $\psi_+^p$ & $-1$ \\
         $\psi_-^i$ & $1$ &
         $\psi_-^p$ & $0$ 
  \end{tabular}
\end{center}
(It is straightforward to check this symmetry is anomaly-free.)
Imposing boundary conditions on the $p$ field will remove its
zero modes except for $p=0$, and so will force the space of bosonic
zero modes to be the zero section of the total space of ${\cal O}(-5)$.
When we restrict to that zero section, there is a short exact sequence
of holomorphic bundles
\begin{displaymath}
0 \: \longrightarrow \: T {\bf P}^4 \: \longrightarrow \:
T {\cal O}(-5) |_{ {\bf P}^4 } \: \longrightarrow \:
{\cal O}(-5) \: \longrightarrow \: 0
\end{displaymath}
and so as smooth bundles,
\begin{displaymath}
T {\cal O}(-5) |_{ {\bf P}^4 } \: \cong \:
T {\bf P}^4 \oplus {\cal O}(-5)
\end{displaymath}
However, the different components of the restriction of the tangent
bundle have different boundary conditions along the time axis.

Let us next figure out what spacelike boundary conditions the Landau-Ginzburg
model fields should possess
so as to RG flow to the NS sector theory.
To that end,
note that if some of the fields have NS boundary conditions, then
interactions in the theory can demand that other fields also have
NS boundary conditions.  After all, the boundary conditions must act
by a symmetry of the theory, else the form of the action will
be coordinate-dependent.  (Put another way, NS and R boundary conditions
arise in the GSO ${\bf Z}_2$ orbifold, but one can only orbifold by
a symmetry.)  In particular, because of the interaction terms
\begin{displaymath}
\psi_+^i \psi_-^p D_i G \: + \: \psi_+^i \psi_-^j p D_i D_j G
\: + \: \psi_+^p \psi_-^i D_i G \: + \: \cdots
\end{displaymath}
we see that if the fermions $\psi_+^i$ have R boundary conditions and
$\psi_-^i$ have NS boundary conditions, then the $p$ field must also
have NS boundary conditions, as too must its superpartner $\psi_+^p$,
though its other superpartners $\psi_-^p$ must have R boundary conditions.
For later use in computing the genus matching the R sector case,
note that
if the fermions $\psi_+^i$ and $\psi_-^i$ both have R boundary
conditions, then so too must $p$ and both its superpartners $\psi^p_{\pm}$.

Let us also work out the action of $(-)^{F_R}$, since it appears in the trace
defining the elliptic genus \cite{edoldeg1}.  
Under the right R-charge, $\phi^i$ and
$\psi_-^i$ have charge $0$, and $\psi_+^i$ charge $1$.
In order to be consistent with interaction terms, as above,
$p$ and $\psi_-^p$ must have charge $-1$, and $\psi_+^p$ charge $0$.

Let us check this analysis by writing out the contributions of the
nonzero right-moving modes (in the sector which flows to the NS theory), 
which should all cancel out.
The right-moving nonzero modes of the $\phi$ and $p$ fields should
contribute a factor of
\begin{displaymath}
\ch\left(
\bigotimes_{n=1,2,3,\cdots} S_{\overline{q}^n}
\left( (T {\bf P}^4 )^ {\bf C} \right)
\bigotimes_{n=1/2,3/2,\cdots} S_{-\overline{q}^n}
\left( z {\cal O}(-5)^{ {\bf C} }
\right)
\right)
\end{displaymath}
(where the $-q^n$ on the $p$ contribution is due to the presence of the
$(-)^{F_R}$ in the trace, as discussed above, and $z = \exp(- i \gamma)$),
and the nonzero modes of $\psi_+^i$, $\psi_+^p$ contribute
a factor of
\begin{displaymath}
\ch\left(
\bigotimes_{n=1,2,3,\cdots} \Lambda_{-\overline{q}^n}\left(
(T {\bf P}^4 )^{\bf C} \right)
\bigotimes_{n=1/2,3/2,\cdots} \Lambda_{\overline{q}^n}\left(
z {\cal O}(-5)^{ {\bf C} }
\right)
\right)
\end{displaymath}
However, it can be shown (see appendix~\ref{app:identities})
that
\begin{displaymath}
S_{q}( {\cal E} ) \: = \: \left( \Lambda_{-q}({\cal E}) \right)^{-1}
\end{displaymath}
and so we see that the total contribution of the right-moving nonzero
modes cancels out.  Note in particular that the factor of $(-)^{F_R}$ in
the trace 
was essential in order
for this to happen.

Putting this together, and using the multiplicative properties of
the $S_{q^n}$ and $\Lambda_{q^n}$,
we get that the elliptic genus of the
Landau-Ginzburg model that RG flows to the NS theory
is given by
\begin{eqnarray*}
\lefteqn{
q^{-(1/24)(-2 + 2(4) + (1)(4) -1)} 
\overline{q}^{-(1/24)( 2(4) -2(4) -1 +1)}
} \\
& & \cdot
\int_{ {\bf P}^4 } \Td( T {\bf P}^4 ) \wedge \ch \left(
\left(  {\cal O}  \ominus  {\cal O}(-5)  \right)
\bigotimes_{n=1,2,3,\cdots}
 S_{q^n}\left( (T {\bf P}^4)^{\bf C} \right) 
\bigotimes_{n=1/2,3/2,\cdots} S_{-q^n } \left( (
z {\cal O}(-5) )^{\bf C}
\right)
\right. 
 \\
& & \hspace*{1.4in} \cdot \left.
\bigotimes_{n=1/2,3/2,\cdots}
 \Lambda_{q^n  } \left( ( z^{-1} T {\bf P}^4 )^{\bf C}
\right)
\bigotimes_{n=1,2,\cdots} \Lambda_{-q^n} \left( ({\cal O}(-5))^{\bf C} 
\right)
\right)
\end{eqnarray*}
The factor of ${\cal O} \ominus {\cal O}(-5)$ arises
from the zero modes of $\psi_-^p$, which is in the R sector, and odd
under $(-)^{F_R}$.
The $S_{q^n}$ factor arises from the $\phi$ modes, and
the $S_{-q^n}$ factor from the $p$ modes.  It has a $-q^n$ instead of
$q^n$ because the $p$ field has $(-)^{F_R} = -1$.  Similarly, the
$\Lambda_{q^n}$ factor comes from the $\psi_-^i$ modes (which have
$(-)^{F_R} = +1$), and the
$\Lambda_{-q^n}$ from the $\psi_-^p$ modes (which have $(-)^{F_R}=-1$).  
The overall factors of $q$, $\overline{q}$ are 
determined by the zero energy of a set of
free $p$, $\psi_{\pm}^p$, $\phi^i$, $\psi_{\pm}^i$ fields, subject to the
boundary conditions discussed previously.

Next, we shall check that the Landau-Ginzburg elliptic genera match
the corresponding nonlinear sigma model elliptic genera, which should
follow physically from the fact that they are in the same universality class.
We will begin with the Landau-Ginzburg elliptic genus that RG flows
to the NS sector nonlinear sigma model elliptic genus.
To do this, we will use results on Thom classes for elliptic genera
which are derived in appendix~\ref{app:thom}.
Using Thom classes, the NS sector elliptic genus of the nonlinear
sigma model on the quintic should match
the integral over ${\bf P}^4$ of
\begin{displaymath}
\Td( T {\bf P}^4 ) \wedge \ch \left(
\bigotimes_{n=1,2,3,\cdots} 
S_{q^n}\left( (T {\bf P}^4)^{\bf C} \right) 
\bigotimes_{n=1/2,3/2,5/2,\cdots} \Lambda_{q^n  } \left( 
(z^{-1} T {\bf P}^4)^{\bf C} \right)
\right)
\end{displaymath}
times the Thom class for the embedding,
which is (appendix~\ref{app:thom:lg})
\begin{displaymath}
\ch\left(
\bigotimes_{n=1/2,3/2,5/2,\cdots} S_{-q^n } \left( 
( z {\cal O}(-5))^{\bf C} \right)
\bigotimes_{n=1,2,3,\cdots} \Lambda_{-q^n} \left( ({\cal O}(-5))^{\bf C} 
\right)
\right)
\end{displaymath}
which exactly matches the physics computation in the Landau-Ginzburg model
above.  Thus, we see explicitly
that in the NS sector,
the elliptic genus of the Landau-Ginzburg model
on the total space of ${\cal O}(-5) \rightarrow {\bf P}^4$ matches
that of the quintic nonlinear
sigma model, as expected.

This form of the Thom class may look rather complicated,
but can be understood relatively easily (and naively) as the
$S^1$-equivariant Thom class on the loop space, following the language
and ideas of \cite{edoldeg2}.

A similar analysis can be done for the R sector genera.
Here, the genus
\begin{displaymath}
{\rm Tr}_{R,R} (-)^{F_R} \exp(i \gamma J_L) q^{L_0}
\overline{q}^{\overline{L}_0}
\end{displaymath}
is given by
\begin{eqnarray}\label{eq:35}
\int_Q \Td (TQ)   %\hat{A}(TX) 
\wedge {\rm ch}\left( 
z^{-3/2} \Lambda_1\left( z T^*Q \right)
\bigotimes_{n=1,2,3,\cdots}
S_{q^n}((TQ)^{\bf C}) 
\bigotimes_{n=1,2,3,\cdots}
 \Lambda_{q^n}\left((z^{-1} TQ)^{\bf C}\right)
\right)
\end{eqnarray}
(using the fact that $\det TQ \cong {\cal O}$).

Next, let us compute the corresponding elliptic genus in the
Landau-Ginzburg model.
We can follow a very similar analysis to the NS sector case.
Half-integral modings become integral modings, and there are
additional contributions from zero modes.
Specifically, from the bosonic $p$ field zero modes, there is a 
contribution
\begin{displaymath}
S_{-1}\left( ( z {\cal O}(-5) )^{\bf C} \right)
\end{displaymath}
from the fermionic $\psi_-^i$ zero modes, a contribution
\begin{displaymath}
z^{+4/2} \Lambda_1( z^{-1} T {\bf P}^4 )
\end{displaymath}
from the fermionic $\psi_-^p$ zero modes, a contribution
\begin{displaymath}
\Lambda_{-1}( {\cal O}(-5) )
\end{displaymath}
%from the fermionic $\psi_+^i$ zero modes, a contribution
%\begin{displaymath}
%\Lambda_{-1}( T {\bf P}^4 )
%\end{displaymath}
and from the fermionic $\psi_+^p$ zero modes, a contribution
\begin{displaymath}
z^{-1/2} \Lambda_1( z {\cal O}(-5) )
\end{displaymath}
Putting this all together, 
we find
\begin{eqnarray}\label{eq:34}
\lefteqn{
q^{0}
} \nonumber \\
& & \cdot \int_{ {\bf P}^4 }
\Td( T {\bf P}^4 ) \wedge \ch\Biggl(
z^{+4/2} z^{-1/2} %\left( \det T^* {\bf P}^4 \right)^{1/2}
\Lambda_1(z^{-1} T {\bf P}^4 )
\otimes
\Lambda_{-1}( {\cal O}(-5) )
%\otimes
%\Lambda_{-1}(T {\bf P}^4 )
\otimes
\Lambda_1( z {\cal O}(-5) )
 \nonumber \\
& & \hspace*{1.5in} \cdot
\bigotimes_{n=1,2,3,\cdots} S_{q^n}( (T {\bf P}^4)^{\bf C} )
\bigotimes_{n=0,1,2,\cdots} S_{-q^n}( ( z {\cal O}(-5) )^{\bf C} )
\nonumber \\
& & \hspace*{1.5in} \left. \cdot
\bigotimes_{n=1,2,3,\cdots} \Lambda_{q^n}( (z^{-1} T {\bf P}^4 )^{\bf C} )
\bigotimes_{n=1,2,3,\cdots} \Lambda_{-q^n}( ({\cal O}(-5) )^{\bf C} )
\right).
\end{eqnarray}
The mathmatical demonstration that \eqref{eq:35} and \eqref{eq:34}
coincide is given in \S\ref{sec:r-sector-genera}.

So far we have discussed only the quintic hypersurface in ${\bf P}^4$,
but the analysis trivially extends to more general complete intersections.
Consider a NLSM on a complete intersection $Y \equiv \{ G_{\mu} = 0 \}
\subset B$ defined by $G_{\mu} \in \Gamma({\cal G})$,
${\cal G}$ a holomorphic vector bundle ${\cal E}$.
Assume $B$ has complex dimension $b$, and $Y$ has complex dimension $y$
(so ${\cal G}$ has rank $b-y$).
(We assume $Y$ is Calabi-Yau.)
The elliptic genus of this nonlinear sigma model, with NS boundary conditions
along spacelike directions on the left-moving fermions, is given by
\begin{equation}\label{eq:14}
q^{-(1/24)(3y)} \int_Y \hat{A}(TY) \wedge {\rm ch}\,\left(
\bigotimes_{n=1,2,3,\cdots} S_{q^n}\left( (TY)^{\bf C} \right) 
\bigotimes_{k=1/2,3/2,5/2,\cdots} \Lambda_{q^k}\left(
(z^{-1} TY )^{\bf C} \right)
\right)
\end{equation}

Corresponding to this nonlinear sigma model is a Landau-Ginzburg model
on
\begin{displaymath}
X \: = \: {\rm Tot}\,\left(
{\cal G}^{\vee} \: \stackrel{\pi}{\longrightarrow} \: B \right)
\end{displaymath}
with superpotential $W = p^{\mu} G_{\mu}$.
It is straightforward to check that the left-moving $U(1)$
symmetry $J_L$ will be anomaly-free so long as\footnote{
We are implicitly using the somewhat obscure fact that if $p$ is
a coordinate on the total space of a bundle ${\cal F}$, say,
then $\psi^p$ zero modes are sections of ${\cal F}^{\vee}$.
}
\begin{displaymath}
\left( \Lambda^{\rm top} TB \right) \otimes
\left( \Lambda^{\rm top} {\cal G}^{\vee} \right)
\: \cong \: K_X^{-1}
\end{displaymath}
is trivializable, hence $X$ is Calabi-Yau.  
(We will get analogous results in other
cases, but from now on will for brevity usually omit this part of the
analysis.)
Following almost exactly the same analysis as for the quintic,
the corresponding elliptic genus of this Landau-Ginzburg model is given by
\begin{eqnarray}\label{eq:38}
\lefteqn{
q^{-(1/24)(2b + b - (b-y) - 2(b-y)}
} \nonumber \\
& & \cdot
\int_B \Td(T B) \wedge {\rm ch}\,\left(
%\left( \bigoplus_{i=0}^{b-y} \left( \ominus^i \Lambda^i {\cal G}^{\vee} \right)
%\right)
\Lambda_{-1}\left( {\cal G}^{\vee} \right)
\bigotimes_{n=1,2,3,\cdots} S_{q^n}\left( (TB)^{\bf C} \right)
\bigotimes_{n=1/2,3/2,\cdots} S_{-q^n}\left( (z {\cal G}^{\vee} )^{\bf C} 
\right)
\right. \nonumber \\
& & \hspace*{1.5in} \cdot \left.
\bigotimes_{n=1/2,3/2,\cdots} \Lambda_{q^n}\left( (z^{-1} TB )^{\bf C} \right)
\bigotimes_{n=1,2,3,\cdots} \Lambda_{-q^n}\left(
( {\cal G}^{\vee} )^{\bf C} \right)
\right)
\label{genus-genlci}
\end{eqnarray}
This matches equation~(\ref{eq:8}) in appendix~\ref{app:thom:lg}, 
using the fact that, for example,
\begin{displaymath}
(z {\cal G}^{\vee})^{\bf C} \: = \:
z {\cal G}^{\vee} \oplus z^{-1} {\cal G} \: = \:
( z^{-1} {\cal G} )^{\bf C}
\end{displaymath}
As discussed in appendix~\ref{app:thom:lg},
this matches equation~(\ref{eq:14}).

Similarly, in the R sector, the nonlinear sigma model genus is
\begin{eqnarray}
\lefteqn{
\int_Y \Td(TY) \wedge \ch\Biggl( 
z^{-y/2} \left( \det TY \right)^{1/2} \Lambda_1( z T^* Y) 
} \nonumber \\
& & \hspace*{1.5in} \cdot \left.
\bigotimes_{n=1,2,3,\cdots} S_{q^n}( (TY)^{\bf C} )
\bigotimes_{n=1,2,3,\cdots} \Lambda_{q^n}( (z^{-1} TY)^{\bf C} )
\right)  
\label{eq:41}
\end{eqnarray}
The corresponding R sector Landau-Ginzburg genus is given by
\begin{eqnarray}\label{eq:37}
\lefteqn{
q^{0}
} \nonumber \\
& & \cdot \int_B \Td(TB) \wedge \ch\Biggl(
z^{+b/2} z^{-(b-y)/2} \Lambda_1( z^{-1} TB) \otimes 
\Lambda_{-1}( {\cal G}^{\vee} ) \otimes
%\Lambda_{-1}( TB ) \otimes
\Lambda_1( z {\cal G}^{\vee} )
\nonumber \\
& & \hspace*{1.5in} \cdot
\left( \det T^*B \right)^{1/2}
\left( \det {\cal G}^{\vee} \right)^{-1/2}
\nonumber \\
& & \hspace*{1.5in} \cdot
\bigotimes_{n=1,2,3,\cdots} S_{q^n}( (TB)^{\bf C} )
\bigotimes_{n=0,1,2,\cdots} S_{-q^n}( (z {\cal G}^{\vee} )^{\bf C} )
\nonumber \\
& & \hspace*{1.5in} \left. \cdot
\bigotimes_{n=1,2,3,\cdots} \Lambda_{q^n}( (z^{-1} TB)^{\bf C} )
\bigotimes_{n=1,2,3,\cdots} \Lambda_{-q^n}( ({\cal G}^{\vee})^{\bf C} )
\right)
\end{eqnarray}
Note that
\begin{displaymath}
K_X \: = \: \left( \det T^*B \right) \otimes
\left( \det {\cal G} \right)
\end{displaymath}
so when $X$ is Calabi-Yau, the middle row of determinant factors is
trivial.   In \S\ref{sec:r-sector-genera} we show that the genera
\eqref{eq:41} and \eqref{eq:37} coincide.

\subsection{Models realizing cokernels of maps}

Beginning here and in the next several subsections, we will
compute elliptic genera of (0,2) Landau-Ginzburg models that
renormalization-group flow to heterotic NLSM's, as described in
\cite{alg02}.

In this subsection we will study
a heterotic Landau-Ginzburg model
that should flow under the renormalization group to a heterotic
NLSM on a space $B$ 
with a bundle ${\cal E}'$ defined as the
cokernel of an injective map:
\begin{displaymath}
{\cal E}' \: = \: \mbox{coker }\left\{
{\cal F}_1 \: \stackrel{\tilde{E}}{\longrightarrow} \: {\cal F}_2 \right\}.
\end{displaymath}
The corresponding heterotic Landau-Ginzburg model will be defined
over the space 
\begin{displaymath}
X \: = \: \mbox{Tot}\left( {\cal F}_1 \: \stackrel{\pi}{\longrightarrow}
\: B \right),
\end{displaymath}
with gauge bundle ${\cal E} \equiv \pi^* {\cal F}_2$,
all $F_a \equiv 0$,
and $E^a = p \tilde{E}^a$ for
$p$ fiber coordinates along ${\cal F}_1$.

The NS sector 
elliptic genus of the NLSM is given by
\begin{equation}  \label{eg:nlsm:cokernel}
q^{-(1/24)(2n+r)}
\int_B \Td(TB) %\hat{A}(B)
\wedge\ch\left(
\bigotimes_{n=1,2,3,\cdots} S_{q^n}\left( (TB)^{\bf C} \right) 
\bigotimes_{n=1/2,3/2,\cdots}
\Lambda_{q^n  }\left( (z^{-1} {\cal E}' )^{\bf C}
\right)
\right)
\end{equation}
where $n$ is the dimension of $B$, and $r$ is the rank of
${\cal E}'$.
We are implicitly assuming that $B$ is a spin manifold (otherwise 
we cannot make sense of the
right-moving R sector vacuum).
This is computed as the one-loop partition function in which the left-moving 
fermions $\lambda_-$ have NS boundary conditions along the spatial direction:
\begin{displaymath}
\lambda_-(x_1+1,x_2) \: = \: - \lambda_-(x_1,x_2),
\end{displaymath}
This is why the $\Lambda_{q^n }$ factors are half-integrally
moded -- reflecting the NS boundary conditions along the spatial axis.

In principle, this should match the corresponding
elliptic genus of the Landau-Ginzburg
model, which we shall describe next.
As before, to define the genus, we must twist the boundary conditions of
the fields by a left-moving $U(1)$ current that commutes
with the right-moving supercharge.  Consistent choices must
leave the interaction terms
\begin{displaymath}
\psi_+^i \lambda_-^{\overline{a}} p D_i \tilde{E}^b h_{\overline{a} b}
\: + \: \psi_+^p \lambda_-^{\overline{a}} \tilde{E}^b h_{\overline{a} b}
\: + \: \mbox{cc}
\end{displaymath}
invariant.
For example, if we give the $\lambda_-^a$ NS boundary conditions along
the spacelike axis, then we must also give $p$ and $\psi_+^p$ NS boundary
conditions along the spacelike axis.
More generally, the fields will have 
charges as follows: 
\begin{center}
  \begin{tabular}{c|c|c|c}
         Field & $Q_L$  & Field & $Q_L$ \\ \hline
         $\phi^i$ & $0$ & $p$ & $+1$ \\
         $\psi_+^i$ & $0$ & $\psi_+^p$ & $+1$ \\
         $\lambda_-^a$ & $+1$ &
          & 
  \end{tabular}
\end{center}
(Note that in order for this symmetry to RG flow to the corresponding
NLSM symmetry, $\lambda_-^a$ must have charge $+1$, which then determines
the other charges.)
Because of the interaction terms above\footnote{
This choice is not unique.  Another choice compatible with the interaction
terms is to take $(-)^{F_R}$ to be $-1$ for $\psi_+^i$, $\psi_+^p$,
and $\lambda_-^a$.  This choice also yields to all $\overline{q}$ contributions
cancelling out, and RG flows to a nonlinear sigma model in which
the trace over states has a 
$(-)^{F_R}(-)^{F_L}$ instead of just $(-)^{F_R}$.
Our choice above is made to reproduce the trace over states given earlier.
}, $(-)^{F_R}$ will be
$-1$ for $\psi_+^i$ and $p$. 
The nontrivial boundary condition on the $p$ field, but not the
$\phi^i$ fields, means that the space of bosonic zero modes will
be the zero section of the vector bundle $X$, {\it i.e.} a copy
of $B$.  Furthermore, as smooth bundles,
\begin{displaymath}
TX|_B \: \cong \: TB \oplus {\cal F}_1
\end{displaymath}
in which the fermions coupling to the two factors will have different
boundary conditions.

Now, we shall compute the NS sector elliptic genus in the Landau-Ginzburg
model, and compare to the corresponding genus in the NLSM.
Using the charges above, we see that
the elliptic genus of the Landau-Ginzburg model is
\begin{eqnarray}   
\lefteqn{
q^{-(1/24)( +2n - r_1  + r_2)}
\overline{q}^{-(1/24)(+2n - r_1 - 2n + r_1)}
} \nonumber \\
& & \cdot
\int_B \Td(TB) %\hat{A}(B) 
\wedge \ch\left(
\bigotimes_{n=1,2,3\cdots} S_{q^n}\left((TB)^{\bf C}\right) 
\bigotimes_{n=1/2,3/2,\cdots} S_{-q^n }\left(
( z^{-1} {\cal F}_1)^{\bf C}\right)
\right.  \nonumber\\
& & \hspace*{2.5in}
\left.
\bigotimes_{n=1/2,3/2,\cdots} \Lambda_{q^n }\left( 
( z^{-1} {\cal F}_2)^{\bf C} \right)
\right)  \label{lgcokernel}
\end{eqnarray}
where the two $S_{q^n}$ factors arise from modes of the $\phi$ and $p$
fields, and the $\Lambda_{q^n}$ factor from modes of $\lambda_-^a$.
(Contributions from $\psi_+^i$ and $\psi_+^p$ are easily checked to
cancel out against right-moving contributions from $\phi^i$ and $p$.)
The $r_i$ are the ranks of the bundles ${\cal F}_i$.

Let us compare this to the elliptic genus in equation~(\ref{eg:nlsm:cokernel}),
using the identities in appendix~\ref{app:identities}.
From the definition of ${\cal E}'$ and the identities in
appendix~\ref{app:identities} we have immediately that
\begin{displaymath}
\Lambda_{q}(z^{-1} {\cal E}') \: = \: 
\Lambda_{q}(z^{-1} {\cal F}_2) \left( \Lambda_{q}(z^{-1} {\cal F}_1) 
\right)^{-1}
\: = \:
\Lambda_{q}(z^{-1} {\cal F}_2)
S_{-q}(z^{-1} {\cal F}_1)
\end{displaymath}
from which it immediately follows that the elliptic genus of the
Landau-Ginzburg model (\ref{lgcokernel}) matches that of the nonlinear
sigma model (\ref{eg:nlsm:cokernel}) to which it flows under the renormalization
group \cite{alg02}, as expected.

Next, we shall work through the corresponding computations for
R sector genera.  The R sector elliptic genus of the NLSM is given by
\begin{eqnarray*}
\lefteqn{
q^{+(1/12)(r_2 - r_1 - n)}
} \nonumber \\
& & \cdot
\int_B \Td(TB) \wedge \ch\Biggl(
z^{-(r_2-r_1)/2} \left( \det {\cal E}' \right)^{1/2}
\Lambda_1( z {\cal E}'^{\vee} )
\nonumber \\
& & \hspace*{1.5in} \left. \cdot
\bigotimes_{n=1,2,3,\cdots} S_{q^n}( (TB)^{\bf C} )
\bigotimes_{n=1,2,3,\cdots} \Lambda_{q^n}( (z^{-1} {\cal E}')^{\bf C} )
\right)
\end{eqnarray*}
The corresponding R sector Landau-Ginzburg genus is given by
\begin{eqnarray}
\lefteqn{ q^{-(1/24)(2n + 2r_1 - 2r_2)} 
} \nonumber \\
& & \cdot \int_B \Td(TB) \wedge \ch\Biggl(
z^{+r_2/2} \Lambda_{1}(z^{-1} {\cal F}_2) z^{+r_1/2} \Lambda_{1}(z^{-1} 
{\cal F}_1)
\nonumber \\
& & \hspace*{1.5in} \cdot
\left( \det {\cal F}_2 \right)^{-1/2} 
\left( \det {\cal F}_1 \right)^{-1/2}
\nonumber \\
& & \hspace*{1.5in} \cdot
\bigotimes_{n=1,2,3,\cdots} S_{q^n}((TB)^{\bf C}) 
\bigotimes_{n=0,1,2,\cdots} S_{-q^n}( (z^{-1} {\cal F}_1)^{\bf C} )
\nonumber \\
& & \hspace*{1.5in} \left. \cdot
\bigotimes_{n=1,2,3,\cdots} \Lambda_{q^n}( (z^{-1} {\cal F}_2)^{\bf C} )
\right)
\end{eqnarray}
where the 
\begin{displaymath}
z^{+r_2/2} \left( \det {\cal F}_2 \right)^{-1/2}
\Lambda_{1}(z^{-1} {\cal F}_2)
\end{displaymath}
factor is from 
$\lambda_-$ zero modes, the 
\begin{displaymath}
z^{+r_1/2} \left( \det {\cal F}_1 \right)^{-1/2}
\Lambda_{1}(z^{-1} {\cal F}_1)
\end{displaymath}
factor from
$\psi_+^p$ zero modes, and the $S_{-1}((z^{-1} {\cal F}_1)^{\bf C} )$
factor from $p$ zero modes.

One can show that the R sector Landau-Ginzburg genus matches that of the
NLSM in the same fashion as before.  The factors from nonzero modes
combine in exactly the same form as before,
and for the zero modes, note that
\begin{eqnarray*}
\lefteqn{
z^{+r_2/2} \left( \det {\cal F}_2 \right)^{-1/2}
\Lambda_{1}(z^{-1} {\cal F}_2)
z^{+r_1/2} \left( \det {\cal F}_1 \right)^{-1/2}
\Lambda_{1}(z^{-1} {\cal F}_1)
S_{-1}\left( z^{-1} {\cal F}_1 \right)
S_{-1} \left( z {\cal F}_1^{\vee} \right)
} \\
&\hspace*{1in} = &
z^{-r_2/2} \left( \det {\cal F}_2 \right)^{1/2}
\Lambda_{1}(z {\cal F}_2^{\vee})
z^{+r_1/2}
 \left( \det {\cal F}_1 \right)^{-1/2}
S_{-1} \left( z {\cal F}_1^{\vee} \right)
\\
& \hspace*{1in} = &
z^{-(r_2-r_1)/2}
\left( \det {\cal E}' \right)^{1/2}
\Lambda_1\left(z {\cal E}'^{\vee} \right)
\end{eqnarray*}
which shows that the R sector Landau-Ginzburg genus matches the
R sector NLSM genus.

\subsection{Models realized as kernels of maps}

Suppose we have a heterotic NLSM on a space $B$
with gauge bundle given by the kernel ${\cal E}'$ of the short
exact sequence
\begin{displaymath}
0 \: \longrightarrow \: {\cal E}' \: \longrightarrow \:
{\cal F}_1 \: \stackrel{F_a}{\longrightarrow} \: {\cal F}_2 \: 
\longrightarrow \: 0.
\end{displaymath}
Applying ideas from \cite{alg02,dk},
this heterotic NLSM should be in the same
universality class as a heterotic Landau-Ginzburg model 
on
\begin{displaymath}
X \: = \: \mbox{Tot}\left( {\cal F}_2^{\vee} \: \stackrel{\pi}{\longrightarrow}
\: B \right),
\end{displaymath}
with gauge bundle ${\cal E} = \pi^* {\cal F}_1$, $E^a \equiv 0$,
and $F_a = p \tilde{F}_a$ 
defined by the map $\tilde{F}_a: {\cal F}_1 \rightarrow {\cal F}_2$
defining ${\cal E}'$ and $p$ fiber coordinates on ${\cal F}_2$.

The NS sector elliptic genus of the NLSM is 
\begin{equation}   \label{eg:nlsm:kernel}
q^{-(1/24)(2n+r)}
\int_B \Td(TB)  %\hat{A}(B)
\wedge\ch\left(\bigotimes_{n=1,2,3,\cdots}
 S_{q^n}\left( (TB)^{\bf C} \right) 
\bigotimes_{n=1/2,3/2,\cdots}
\Lambda_{q^n }\left( ( z^{-1} {\cal E}')^{\bf C} 
\right)
\right)
\end{equation}
where $n$ is the dimension of $B$ and $r$ is the rank of ${\cal E}'$.
As before, we are implicitly assuming that $B$ is a spin manifold.

Next, let us compute the elliptic genus of the Landau-Ginzburg model.
As before, we must twist by a left-moving $U(1)$ symmetry, which is
determined by the interactions
\begin{displaymath}
\psi_+^i \lambda_-^a p D_i \tilde{F}_a \: + \:
\psi_+^p \lambda_-^a \tilde{F}_a \: + \: \mbox{cc}
\end{displaymath}
This determines the charges to be
\begin{center}
  \begin{tabular}{c|c|c|c}
         Field & $Q_L$  & Field & $Q_L$ \\ \hline
         $\phi^i$ & $0$ & $p$ & $-1$ \\
         $\psi_+^i$ & $0$ & $\psi_+^p$ & $-1$ \\
         $\lambda_-^a$ & $+1$ &
          &
  \end{tabular}
\end{center}
(almost, but not quite, the same as for the previous example of a cokernel).
In addition, because of the interaction terms, if we put NS boundary
conditions on the $\lambda_-^a$ along spacelike directions, then we must also
put NS boundary conditions on $p$ and $\psi_+^p$.
For the same reason, $(-)^{F_R}$ must have value $-1$ for\footnote{
As in the last section, this choice is ambiguous; we make the choice
that RG flows to the NLSM elliptic genus given earlier, with the
state trace containing $(-)^{F_R}$ not $(-)^{F_R + F_L}$.
} 
$\psi_+^i$ and $p$.
The resulting elliptic genus of the Landau-Ginzburg model is 
\begin{eqnarray}   
\lefteqn{
q^{-(1/24)(2n - r_2 + r_1)}
\overline{q}^{-(1/24)(2n - r_2 -2n + r_2)}
} \nonumber \\
& & \cdot
\int_B \Td(TB)   %\hat{A}(B) 
\wedge \ch\left(
\bigotimes_{n=1,2,3,\cdots} S_{q^n}\left( (TB)^{\bf C}
\right) 
\bigotimes_{n=1/2,3/2,\cdots} S_{-q^n }
\left( ( z {\cal F}_2^{\vee})^{\bf C} \right)
\right.  \nonumber \\
& & \hspace*{2.5in} \left.
\bigotimes_{n=1/2,3/2,\cdots} \Lambda_{q^n }
\left( ( z^{-1} {\cal F}_1)^{\bf C} \right)
\right) \label{lgkernel}
\end{eqnarray}
where the first $S_{q^n}$ factor is from modes of the $\phi^i$ field,
the second from modes of the $p$ field, and the $\Lambda_{-q^n}$ factor
is from modes of the $\lambda_-^a$ field.
The $r_i$ are the ranks of the ${\cal F}_i$, and the overall $q$ and
$\overline{q}$ factors are determined by the zero energy contributions
of the fields.

Finally, let us compare these two elliptic genera.
From the definition of ${\cal E}'$ and the identities in 
appendix~\ref{app:identities}, we have that
\begin{displaymath}
\Lambda_{q}(z {\cal E}') \: = \:
\Lambda_{q}(z {\cal F}_1) \left( \Lambda_{q}(z {\cal F}_2) \right)^{-1}
\: = \:
\Lambda_{q}(z {\cal F}_1) S_{-q}(z {\cal F}_2)
\end{displaymath}
so the elliptic genera of the two representatives of the same
universality class match, as expected.

Next, we shall work through the corresponding computations for R sector
genera.  The R sector elliptic genus of the NLSM is given by
\begin{eqnarray*}
\lefteqn{
q^{+(1/12)(r_1 - r_2 - n)}
} \nonumber \\
& & \cdot
\int_B \Td(TB) \wedge \ch\Biggl(
z^{-(r_1-r_2)/2} \left( \det {\cal E}' \right)^{1/2}
\Lambda_1( z {\cal E}'^{\vee} )
\nonumber \\
& & \hspace*{1.5in} \left. \cdot
\bigotimes_{n=1,2,3,\cdots} S_{q^n}( (TB)^{\bf C} )
\bigotimes_{n=1,2,3,\cdots} \Lambda_{q^n}( (z^{-1} {\cal E}')^{\bf C} )
\right)
\end{eqnarray*}
The corresponding R sector Landau-Ginzburg genus is given by
\begin{eqnarray}
\lefteqn{
q^{-(1/24)(2n + 2r_2 - 2r_1)}
} \nonumber \\
& & \cdot \int_B \Td(TB) \wedge \ch\Biggl(
z^{+r_1/2} \Lambda_1( z^{-1} {\cal F}_1 )
z^{-r_2/2} \Lambda_1( z {\cal F}_2^{\vee} )
\nonumber \\
& & \hspace*{1.5in} \cdot
\left( \det {\cal F}_1 \right)^{-1/2}
\left( \det {\cal F}_2 \right)^{1/2}
\nonumber \\
& & \hspace*{1.5in} \cdot
\bigotimes_{n=1,2,3,\cdots} S_{q^n}( (TB)^{\bf C} )
\bigotimes_{n=0,1,2,\cdots} S_{-q^n}( (z {\cal F}_2^{\vee})^{\bf C} )
\nonumber \\
& & \hspace*{1.5in} \left. \cdot
\bigotimes_{n=1,2,3,\cdots} \Lambda_{q^n}( (z^{-1} {\cal F}_1)^{\bf C} )
\right)
\end{eqnarray}
We can pair up the nonzero modes to match the R sector NLSM genus in the
same fashion as for the NS sector genera.
The zero modes are related as follows:
\begin{eqnarray*}
\lefteqn{
z^{+r_1/2} \Lambda_1( z^{-1} {\cal F}_1 )
\left( \det {\cal F}_1 \right)^{-1/2}
z^{-r_2/2} \Lambda_1( z {\cal F}_2^{\vee} )
\left( \det {\cal F}_2 \right)^{1/2}
S_{-1}( z {\cal F}_2^{\vee} )
S_{-1}( z^{-1} {\cal F}_2 )
} \\
& \hspace*{1in} = &
z^{-(r_2-r_1)/2}
 \Lambda_1( z^{-1} {\cal F}_1 )
S_{-1}( z^{-1} {\cal F}_2 )
\left( \det {\cal F}_1 \right)^{-1/2}
\left( \det {\cal F}_2 \right)^{1/2}
\\
& \hspace*{1in} = &
z^{-(r_2-r_1)/2}
\Lambda_1( z^{-1} {\cal E}' )
\left( \det {\cal E}' \right)^{-1/2}
\\
& \hspace*{1in} = &
z^{-(r_1-r_2)/2}
\Lambda_1( z {\cal E}'^{\vee} )
\left( \det {\cal E}' \right)^{1/2}
\end{eqnarray*}
Thus, we see that the R sector Landau-Ginzburg genus does indeed
match the R sector NLSM genus, as predicted by renormalization
group flow.

\subsection{Models realized as cohomologies of monads}

Suppose we have a heterotic NLSM on a space $B$
with gauge bundle ${\cal E}'$ given by the cohomology of the short complex
\begin{displaymath}
0 \: \longrightarrow \: {\cal F}_0 \: \stackrel{\tilde{E}^a}{\longrightarrow} \:
{\cal F}_1 \: \stackrel{\tilde{F}_a}{\longrightarrow} \: {\cal F}_2 \:
\longrightarrow \: 0
\end{displaymath}
at the middle term.
Judging from related examples and standard analyses in previous sections
here and in
\cite{alg02,dk}, this heterotic NLSM should be in the same
universality class as a heterotic Landau-Ginzburg model 
on
\begin{displaymath}
X \: = \: \mbox{Tot}\left( 
{\cal F}_0 \oplus {\cal F}_2^{\vee} \: \stackrel{\pi}{\longrightarrow} \:
B \right),
\end{displaymath}
with ${\cal E} \equiv \pi^*{\cal F}_1$,
and $E^a = p' \tilde{E}^a$, $F_a = p \tilde{F}_a$, 
where $p$ are fiber coordinates along ${\cal F}_2^{\vee}$ and $p'$ fiber
coordinates along ${\cal F}_0$.

The elliptic genus of the NLSM is 
\begin{equation}   \label{eg:nlsm:monad}
q^{-(1/24)(2n + r)}
\int_B \Td(TB)  %\hat{A}(B)
\wedge\ch\left(\bigotimes_{n=1,2,3,\cdots} 
S_{q^n}\left( (TB)^{\bf C} \right) 
\bigotimes_{n=1/2,3/2,\cdots}
\Lambda_{q^n }\left( 
( z^{-1} {\cal E}')^{\bf C} \right)
\right)
\end{equation}
where $n$ is the dimension of $B$ and $r$ is the rank of ${\cal E}'$.
As before, we are assuming that $B$ is a spin manifold.

Next, let us compute the elliptic genus of the Landau-Ginzburg model.
As before, we must twist by a left-moving $U(1)$ symmetry, which is
determined by the interaction terms
\begin{displaymath}
\psi_+^i \lambda_-^a p D_i \tilde{F}_a \: + \:
\psi_+^p \lambda_-^a \tilde{F}_a \: + \:
\psi_+^i \lambda_-^{\overline{a}} p' D_i \tilde{E}^b h_{\overline{a} b}
\: + \:
\psi_+^{p'} \lambda_-^{\overline{a}}  \tilde{E}^b h_{\overline{a} b}
\: + \: \mbox{cc}
\end{displaymath}
and is given by the charges
\begin{center}
  \begin{tabular}{c|c|c|c}
         Field & $Q_L$  & Field & $Q_L$ \\ \hline
         $\phi^i$ & $0$ & $\psi_+^i$ & $0$ \\
         $p$ & $-1$ & $\psi_+^p$ & $-1$ \\
         $p'$ & $+1$ & $\psi_+^{p'}$ & $+1$ \\
         $\lambda_-^a$ & $+1$ &
          &
  \end{tabular}
\end{center}
From the interaction terms, the fields $\lambda_-^a$, $p$, $p'$, $\psi_+^p$,
and $\psi_+^{p'}$ will be in the NS sector, where $\phi^i$, $\psi_+^i$ will
be in the R sector.
Furthermore, $(-)^{F_R}$ acts by\footnote{
As before, there is an ambiguity, and we make the choice that flows to
standard conventions in the IR.
} $-1$ on $\psi_+^i$, $p$, and $p'$.

The resulting NS sector
elliptic genus of the Landau-Ginzburg model is 
\begin{eqnarray*}
\lefteqn{
q^{-(1/24)(2n - r_2 - r_0 + r_1) }
\overline{q}^{-(1/24)(2n - r_2 - r_0 -2n + r_2 + r_0)}
} \\
& & \cdot
\int_B \Td(TB)   %\hat{A}(B) 
\wedge \ch \left(
\bigotimes_{n=1,2,3,\cdots} S_{q^n}\left(
(TB)^{\bf C}\right)  
\right.  \\
& & \hspace*{1in} 
\bigotimes_{n=1/2,3/2,\cdots} S_{-q^n }
\left( ( z {\cal F}_2^{\vee})^{\bf C} \right)
\bigotimes_{n=1/2,3/2,\cdots} S_{-q^n }
\left( ( z^{-1} {\cal F}_0)^{\bf C} \right)
\\
& & \hspace*{1in} \left.
\bigotimes_{n=1/2,3/2,\cdots} \Lambda_{q^n }
\left( ( z^{-1} {\cal F}_1)^{\bf C} \right)
\right)
\end{eqnarray*}
where the $S_{q^n}$ factors come from $\phi$, $p$, and $p'$ left-moving modes,
and the $\lambda_{-q^n}$ from $\lambda_-^a$ modes.
The $r_i$ are the ranks of the ${\cal F}_i$.

Let us compare the NS elliptic genus in the Landau-Ginzburg model above
to that of the nonlinear sigma model, in equation~(\ref{eg:nlsm:monad}).
Since these two theories are in the same universality class \cite{alg02},
the elliptic genera should match.
Now, from the definition of ${\cal E}'$ and the identities in 
appendix~\ref{app:identities}, we have that
\begin{displaymath}
\Lambda_{q}(z {\cal E}) \: = \:
\Lambda_{q}(z {\cal F}_1) \left( \Lambda_{q}(z {\cal F}_0) \right)^{-1}
\left( \Lambda_{q}(z {\cal F}_2) \right)^{-1}
\: = \:
\Lambda_{q}(z {\cal F}_1)
S_{-q}(z {\cal F}_0) 
S_{-q}(z {\cal F}_2)
\end{displaymath}
from which we see that, as expected, the elliptic genera match.

Next, we shall work through the corresponding computations for
R sector genera.
The R sector elliptic genus of the NLSM is given by
\begin{eqnarray*}
\lefteqn{
q^{+(1/12)(r_1 - r_0 - r_2 - n)}
} \nonumber \\
& & \cdot
\int_B \Td(TB) \wedge \ch\Biggl(
z^{-(r_1-r_0 - r_2)/2} \left( \det {\cal E}' \right)^{1/2}
\Lambda_1( z {\cal E}'^{\vee} )
\nonumber \\
& & \hspace*{1.5in} \left. \cdot
\bigotimes_{n=1,2,3,\cdots} S_{q^n}( (TB)^{\bf C} )
\bigotimes_{n=1,2,3,\cdots} \Lambda_{q^n}( (z^{-1} {\cal E}')^{\bf C} )
\right)
\end{eqnarray*}
The corresponding R sector Landau-Ginzburg genus is given by
\begin{eqnarray}
\lefteqn{
q^{-(1/24)(2n + 2r_0 + 2r_2 - 2r_1)}
} \nonumber \\
& & \cdot \int_B \Td(TB) \wedge \ch\Biggl(
z^{+r_1/2} \Lambda_1( z^{-1} {\cal F}_1 )
z^{+r_0/2} \Lambda_1( z^{-1} {\cal F}_0 )
z^{-r_2/2} \Lambda_1( z {\cal F}_2^{\vee} )
\nonumber \\
& & \hspace*{1.5in} \cdot
\left( \det {\cal F}_1 \right)^{-1/2}
\left( \det {\cal F}_0 \right)^{-1/2}
\left( \det {\cal F}_2 \right)^{+1/2}
\nonumber \\
& & \hspace*{1.5in} \cdot
\bigotimes_{n=1,2,3,\cdots} S_{q^n}( (TB)^{\bf C} )
\bigotimes_{n=0,1,2,\cdots} S_{-q^n}( (z^{-1} {\cal F}_0)^{\bf C} )
\nonumber \\
& & \hspace*{1.5in} \left. \cdot
\bigotimes_{n=0,1,2,\cdots} S_{-q^n}( (z {\cal F}_2^{\vee} )^{\bf C} )
\bigotimes_{n=1,2,3,\cdots} \Lambda_{q^n}( (z^{-1} {\cal F}_1 )^{\bf C} )
\right)
\end{eqnarray}

We can pair up the nonzero modes to match the R sector NLSM genus in the
same fashion as for the NS sector genera.  The zero modes are related
as follows:
\begin{eqnarray*}
\lefteqn{
z^{+r_1/2} \Lambda_1( z^{-1} {\cal F}_1 )
\left( \det {\cal F}_1 \right)^{-1/2}
z^{-r_0/2} \Lambda_1( z {\cal F}_0^{\vee} )
\left( \det {\cal F}_0 \right)^{1/2}
z^{-r_2/2} \Lambda_1( z {\cal F}_2^{\vee} )
\left( \det {\cal F}_2 \right)^{+1/2}
} \\
& & \hspace*{0.5in} \cdot
S_{-1}( z^{-1} {\cal F}_0 )  S_{-1}( z {\cal F}_0^{\vee} )
S_{-1}( z {\cal F}_2^{\vee} ) S_{-1}( z^{-1} {\cal F}_2 )
 \\
& \hspace*{1in} = &
z^{+r_1/2} \Lambda_1( z^{-1} {\cal F}_1 )
\left( \det {\cal F}_1 \right)^{-1/2}
z^{-r_0/2} \left( \det {\cal F}_0 \right)^{1/2}
\\
& & \hspace*{0.5in} \cdot
z^{-r_2/2} \left( \det {\cal F}_2 \right)^{+1/2}
S_{-1}( z^{-1} {\cal F}_0 )
S_{-1}( z^{-1} {\cal F}_2 )
\\
& \hspace*{1in} = &
z^{(r_1 - r_0 - r_2)/2} 
\left( \det {\cal E}' \right)^{-1/2}
\Lambda_1( z^{-1} {\cal E}')
\\
& \hspace*{1in} = &
z^{-(r_1 - r_0 - r_2)/2}
\left( \det {\cal E}' \right)^{+1/2}
\Lambda_1( z {\cal E}'^{\vee} )
\end{eqnarray*}
Thus, we see that the R sector Landau-Ginzburg genus does indeed match
the R sector NLSM genus, as predicted by renormalization group flow.

\subsection{Models realized as cohomologies of monads over complete
intersections}
\label{sec:models-coh-of-monads}

Here, we consider the most general case.
Suppose we want a heterotic Landau-Ginzburg model that will flow to
a heterotic NLSM on a complete intersection
$Y \equiv \{ G_{\mu} = 0 \} \subset B$ 
defined by $G_{\mu} \in \Gamma({\cal G})$,
${\cal G}$ a holomorphic vector bundle on $B$,
with a gauge bundle ${\cal E}'$ given by the cohomology of the complex of 
holomorphic vector bundles
\begin{displaymath}
0 \: \longrightarrow \: {\cal F}_0|_Y \: \stackrel{\tilde{E}^a|_Y}{
 \longrightarrow} \:
{\cal F}_1|_Y \: \stackrel{\tilde{F}_a|_Y}{\longrightarrow} 
\: {\cal F}_2|_Y \: \longrightarrow \: 0,
\end{displaymath}
where $\tilde{E}_a: {\cal F}_0 \rightarrow {\cal F}_1$
and $\tilde{F}_a: {\cal F}_1 \rightarrow {\cal F}_2$ are defined over
all of $B$, but the sequence above only necessarily becomes a complex 
over $Y \subset B$.  (Furthermore, the complex is exact
everywhere except at the ${\cal F}_1$ term.)  Explicitly, 
\begin{eqnarray*}
 \Ker \tilde{E}^{a} & = & 0, \\
\Coker \tilde{F}_{a} & = & 0, \\
{\cal E}' & = &  \Ker \tilde{F}_{a}/\Img
   \tilde{E}^{a}.
\end{eqnarray*}

Generalizing the (0,2) GLSM description in \cite{alg02,dk},
the corresponding Landau-Ginzburg model is defined over the space
\begin{displaymath}
X \: = \: \mbox{Tot}\left( {\cal F}_0 \oplus
{\cal F}_2^{\vee} \: \stackrel{\pi}{\longrightarrow} \: B
\right),
\end{displaymath}
with gauge bundle ${\cal E}$ an extension\footnote{
In general, the extension will be nontrivial, as an example we will
discuss momentarily will make clear.  Aside from that, we have not 
found a way to uniquely determine the extension in terms of
data of the IR NLSM.
In fact, since renormalization group flow is a lossy process, it is
not completely clear that the Landau-Ginzburg model should be
uniquely determined by the NLSM -- perhaps several
different extensions defining different ${\cal E}$'s in the Landau-Ginzburg
model all flow to the same NLSM.  We have no such examples,
but neither can we rule out the possibility.
} of $\pi^* {\cal F}_1$ by
$\pi^* {\cal G}^{\vee}$:
\begin{displaymath}
0 \: \longrightarrow \: \pi^* {\cal G}^{\vee} \: \longrightarrow \:
{\cal E} \: \longrightarrow \: \pi^* {\cal F}_1 \: \longrightarrow
\: 0.
\end{displaymath}
The $F_a \in \Gamma({\cal E}^{\vee})$ are partially determined by
$G \in \Gamma({\cal G})$ and
\begin{displaymath}
F_a |_{ \pi^* {\cal F}_1^{\vee} } \: = \: p \tilde{F}_a,
\end{displaymath}
where $p$ are fiber coordinates on ${\cal F}_2^{\vee}$
and $\tilde{F}_a$ is the map ${\cal F}_1 \rightarrow {\cal F}_2$.
The $E^a \in \Gamma({\cal E})$ are partially determined by $p' \tilde{E}^a$,
where $p'$ are fiber coordinates on ${\cal F}_0$ and
$\tilde{E}^a$ is the map ${\cal F}_0 \rightarrow {\cal F}_1$.

The NS sector elliptic genus of the NLSM is 
\begin{equation}   \label{eg:nlsm:monad:ci}
q^{-(1/24)(2n + r)}
\int_Y \Td(TY)    %\hat{A}(Y)
\wedge\ch\left(
\bigotimes_{n=1,2,3,\cdots} S_{q^n}\left( (TY)^{\bf C} \right) 
\bigotimes_{n=1/2,3/2,\cdots}
\Lambda_{q^{n}}\left( ( z^{-1} {\cal E}')^{\bf C} 
\right)
\right)
\end{equation}
where $n$ is the dimension of $Y$ and $r$ is the rank of ${\cal E}'$.
As before, we assume $B$ is a spin manifold.

Next, let us compute the elliptic genus of the Landau-Ginzburg model.
As before, we must twist by a left-moving $U(1)$ symmetry, and the charges
under that symmetry are determined in part by interaction terms.
We need to distinguish $\lambda_-$ coupling to $\pi^* {\cal F}_1$
from $\pi^* {\cal G}^{\vee}$; we shall, schematically (ignoring for
the moment the extension), denote the former by $\lambda_-^a$ and
the latter by $\lambda_-^{\alpha}$.  Then, equally schematically,
the interactions will be of the form
\begin{displaymath}
\psi_+^i \lambda_-^a p D_i \tilde{F}_a, \: \: \:
\psi_+^i \lambda_-^{\alpha} D_i G_{\alpha}, \: \: \:
\psi_+^i \lambda_-^{\overline{a}} p' \left( D_i \tilde{E}^b \right) h_{
\overline{a} b}
\end{displaymath}
From these, we can see that if $\phi^i$, $\psi_+^i$ are neutral under
$J_L$, and $\lambda_-^a$ has charge $+1$, then $p$ must have charge $-1$,
in order for the interactions above to remain neutral.  To be a left
R-symmetry, all $\psi_+$'s must have the same left charge as the
corresponding scalars.  Proceeding in this fashion, we find that
the charges determining that left $U(1)$ symmetry are 
\begin{center}
  \begin{tabular}{c|c|c|c}
         Field & $Q_L$  & Field & $Q_L$ \\ \hline
         $\phi^i$ & $0$ & $\psi_+^i$ & $0$ \\
         $p$ & $-1$ & $\psi_+^p$ & $-1$ \\
         $p'$ & $+1$ & $\psi_+^{p'}$ & $+1$ \\
         $\lambda_-^a$ & $+1$ &
          & \\
         $\lambda_-^{\alpha}$ & $0$ & & 
  \end{tabular}
\end{center}

It is straightforward to check that this symmetry is anomaly-free 
so long as
\begin{displaymath}
\left( \Lambda^{\rm top} {\cal F}_1 \right) \otimes
\left( \Lambda^{\rm top} {\cal F}_2 \right)^{\vee} \otimes
\left( \Lambda^{\rm top} {\cal F}_0^{\vee} \right)
\end{displaymath}
is trivializable, which implies that $\Lambda^{\rm top} {\cal E}'$
is trivializable, and the Calabi-Yau condition.

We also 
see that $\phi^i$, $\psi_+^i$, and $\lambda_-^{\alpha}$ are in the R sector,
and $\lambda_-^a$, $p$, $p'$, $\psi_+^p$, $\psi_+^{p'}$ are in the NS sector.
Furthermore, as before, $(-)^{F_R}$ should act by $-1$ 
on $\psi_+^i$, $p$, $p'$, and $\lambda_-^{\alpha}$.

The resulting elliptic genus of the Landau-Ginzburg model is 
\begin{eqnarray}\label{eq:39}
\lefteqn{
q^{-(1/24)(2m - r_2 - r_0 -2 s + r_1)}
\overline{q}^{-(1/24)(2m - r_2 - r_0 - 2m + r_2 + r_0)}
} \nonumber \\
& & \cdot 
\int_B \Td(B) \wedge \ch \left(
\left( {\cal S}_+({\cal G}^{\vee}) \ominus {\cal S}_-({\cal G}^{\vee}) \right)
\bigotimes_{n=1,2,3,\cdots} S_{q^n}\left( (TB)^{\bf C} \right) 
\right.  \nonumber \\
& & \hspace*{1in} 
\bigotimes_{n=1/2,3/2,\cdots} S_{-q^n }
\left( ( z {\cal F}_2^{\vee})^{\bf C} \right)
\bigotimes_{n=1/2,3/2,\cdots} S_{-q^n }
\left( ( z^{-1} {\cal F}_0)^{\bf C} \right)
\nonumber \\
& & \hspace*{1in} \left.
\bigotimes_{n=1/2,3/2,\cdots} \Lambda_{q^n }
\left( ( z^{-1} {\cal F}_1)^{\bf C} \right)
\bigotimes_{n=1,2,3,\cdots} \Lambda_{-q^n}
\left( ( {\cal G}^{\vee})^{\bf C} \right)
\right)    \label{genus-genlcimonad}
\end{eqnarray}
where $m$ is the dimension of $B$, $s$ the rank of ${\cal G}$,
and $r_i$ the rank of ${\cal F}_i$.

Let us compare the NS sector
Landau-Ginzburg elliptic genus above to the NS
sector elliptic genus of the nonlinear sigma model
(\ref{eg:nlsm:monad:ci}).  Since the two theories are in the same
universality class, the two genera ought to match.  In
appendix~\ref{app:thom:vb} we show mathematically that they do match,
as expected.

As a consistency check, let us explicitly describe how to recover
the results on (2,2) complete intersections in section~\ref{22quintic} 
from the expression above.
The (2,2) locus is described by taking
\cite{alg02}
${\cal F}_0 = 0$, ${\cal F}_2 = {\cal G}$, and ${\cal F}_1 = TB$,
so that ${\cal E} = TX$.
It is easy to check that in this case, the expression above reduces to
equation~(\ref{genus-genlci}), as expected.

Next, we shall work through the corresponding computations for
R sector genera.  The R sector elliptic genus of the NLSM is given by
\begin{eqnarray*}
\lefteqn{
q^{+(1/12)(r_1 - r_0 - r_2 - (m-s))}
} \nonumber \\
& & \cdot
\int_Y \Td(TY) \wedge \ch\Biggl(
z^{-(r_1-r_0 - r_2)/2} \left( \det {\cal E}' \right)^{1/2}
\Lambda_1( z {\cal E}'^{\vee} )
\nonumber \\
& & \hspace*{1.5in} \left. \cdot
\bigotimes_{n=1,2,3,\cdots} S_{q^n}( (TY)^{\bf C} )
\bigotimes_{n=1,2,3,\cdots} \Lambda_{q^n}( (z^{-1} {\cal E}')^{\bf C} )
\right)
\end{eqnarray*}
The corresponding R sector Landau-Ginzburg genus is given by
\begin{eqnarray} \label{eq:40}
\lefteqn{
q^{-(1/24)(2m + 2r_0 + 2r_2 - 2s - 2r_1)}
} \nonumber \\
& & \cdot \int_B \Td(TB) \wedge \ch\Biggl(
z^{+ r_1/2} \Lambda_1( z^{-1} {\cal F}_1)
\Lambda_{-1}( {\cal G}^{\vee} )
z^{+r_0/2} \Lambda_1(z^{-1} {\cal F}_0 )
z^{-r_2/2} \Lambda_1( z {\cal F}_2^{\vee} )
\nonumber \\
& & \hspace*{1.5in} \cdot
\left( \det {\cal F}_1 \right)^{-1/2}
\left( \det {\cal F}_0 \right)^{-1/2}
\left( \det {\cal F}_2 \right)^{+1/2}
\nonumber \\
& & \hspace*{1.5in} \cdot
\bigotimes_{n=1,2,3,\cdots} S_{q^n}( (TB)^{\bf C} )
\bigotimes_{n=0,1,2,\cdots} S_{-q^n}( (z^{-1} {\cal F}_0 )^{\bf C} )
\nonumber \\
& & \hspace*{1.5in} \cdot
\bigotimes_{n=0,1,2,\cdots} S_{-q^n}( (z {\cal F}_2^{\vee})^{\bf C} )
\bigotimes_{n=1,2,3,\cdots} \Lambda_{q^n}( (z^{-1} {\cal F}_1 )^{\bf C} )
\nonumber \\
& & \hspace*{1.5in} \left. \cdot
\bigotimes_{n=1,2,3,\cdots} \Lambda_{-q^n}( ({\cal G}^{\vee})^{\bf C} )
\right)
\end{eqnarray}

It is not difficult to adapt the methods of Appendix \ref{app:thom} to
prove that the genera \eqref{eq:39} and \eqref{eq:40} coincide.

\section{General remarks on Thom classes}  \label{thom-genl}

In this paper we have seen that for Landau-Ginzburg models in the
same universality class as nonlinear sigma models,
the elliptic genera of the Landau-Ginzburg models match those of the
nonlinear sigma models via a Thom class computation -- the two expressions
look very different, but the Landau-Ginzburg model computation has the
form of an integral over a larger space of something that localizes
onto the smaller space over which the nonlinear sigma model genus is
defined.

This particular property is not specific to elliptic genera,
but crops up in other contexts as well.
For example, in \cite{alg22}, A-twisted Landau-Ginzburg models were
discussed, and it was observed there that correlation functions in
A-twisted Landau-Ginzburg models in the same universality class
as nonlinear sigma models, matched by virtue of a Thom form,
represented specifically by a Mathai-Quillen form.

\section{Conclusions}

In this paper, we have discussed elliptic genera of both (2,2) and
(0,2) supersymmetric Landau-Ginzburg
models over nontrivial spaces, generalizing methods of \cite{edlg}
for Landau-Ginzburg models over vector spaces.
We checked our results using the renormalization group:
Landau-Ginzburg models in the same universality class as ordinary
nonlinear sigma models should have matching elliptic genera,
which we were able to confirm explicitly.  In those computations,
just as in the A-twisted Landau-Ginzburg model computations of
\cite{alg22,alg02}, we saw that the renormalization group is realized
mathematically via Thom classes.

One direction for future work lies in understanding elliptic genera
and other properties of Landau-Ginzburg models over nontrivial 
{\it stacks} \cite{msx}, in addition to nontrivial spaces.  
One application would be to complete our knowledge of elliptic genera
at various limits of gauged linear sigma models, as `typical' limits
of K\"ahler moduli space are ``hybrid Landau-Ginzburg models,''
which are precisely Landau-Ginzburg models over stacks.
Another application would be to compute elliptic genera of
noncommutative spaces (in the sense of Kontsevich and others,
as opposed to \cite{sw}), as described in \cite{cdhps}
(where they appeared as part of a general discussion of
novel geometries and non-birational phases of abelian GLSM's,
realizing Kuznetsov's ``homological projective duality'' \cite{kuz1})
Those examples of new CFT's
are realized as IR limits of certain hybrid Landau-Ginzburg
models appearing in GLSMs, hence, one way to compute their elliptic genera
would be to compute the elliptic genus of a corresponding Landau-Ginzburg
model on a stack.

\section{Acknowledgements}

E.S. would like to thank J.~Distler for useful conversations.
M.A. and E.S. were partially supported by NSF grant DMS-0705381,
and NSF grant PHY-0755614.

\appendix

\section{Some useful identities}   \label{app:identities}

Define
\begin{eqnarray*}
S_q \left( z{\cal E} \right) & = & 1 \: \oplus \: z q {\cal E} \: \oplus \:
z^2 q^2 \mbox{Sym}^2 {\cal E} \: \oplus \: 
z^3 q^3 \mbox{Sym}^3 {\cal E} \: \oplus \: \cdots \\
& = & S_{zq} {\cal E} \\
\Lambda_q \left( z {\cal E} \right) & = & 1 \: \oplus \: z q {\cal E} 
\: \oplus \:
z^2 q^2 \mbox{Alt}^2 {\cal E} \: \oplus \: 
z^3 q^3 \mbox{Alt}^3 {\cal E} \: \oplus \: \cdots \\
& = & \Lambda_{zq} {\cal E}
\end{eqnarray*}
and similarly
\begin{eqnarray*}
S_q \left( z {\cal E} \right)^{ {\bf C} } & = &
S_q \left( z {\cal E}\right) \otimes S_q \left(
\overline{z} \overline{ {\cal E} } \right) \\
\Lambda_q \left( z {\cal E}\right)^{ {\bf C} } & = &
\Lambda_q \left( z{\cal E} \right) \otimes \Lambda_q 
\left( \overline{z} \overline{ {\cal E} } \right)
\end{eqnarray*}
These expressions should be understood as elements of K-theory of the
underlying space.

These expressions have good multiplicative properties:
\begin{eqnarray*}
S_q \left( {\cal E} \oplus {\cal F} \right)  & = &
\left( S_q {\cal E} \right) \otimes \left( S_q {\cal F} \right) \\
S_q \left( {\cal E} \ominus {\cal F} \right) & = &
\left( S_q {\cal E} \right) \otimes \left(
S_q {\cal F} \right)^{-1} \\
\Lambda_q \left( {\cal E} \oplus {\cal F} \right) & = &
\left( \Lambda_q {\cal E} \right) \otimes
\left( \Lambda_q {\cal F} \right) \\
\Lambda_q \left( {\cal E} \ominus {\cal F} \right) & = &
\left( \Lambda_q {\cal E} \right) \otimes
\left( \Lambda_q {\cal F} \right)^{-1}
\end{eqnarray*}
where we have used the facts that
\begin{eqnarray*}
\mbox{Sym}^n ({\cal E} \oplus {\cal F}) & = &
\bigoplus_{i=0}^n \, \mbox{Sym}^i({\cal E}) \otimes
\mbox{Sym}^{n-i}({\cal F}) \\
\mbox{Alt}^n ({\cal E} \oplus {\cal F}) & = &
\bigoplus_{i=0}^n \, \mbox{Alt}^i({\cal E}) \otimes
\mbox{Alt}^{n-i}({\cal F})
\end{eqnarray*}

Furthermore, the inverses are straightforward to compute.
Using the multiplicative properties above and the splitting principle,
it suffices to consider the action on line bundles.
For a line bundle ${\cal L}$,
\begin{displaymath}
S_q {\cal L} \: = \: 1 \: \oplus \: q {\cal L} \: \oplus \: q^2 {\cal L}^2
\: \oplus \: \cdots \: = \: \frac{1}{1 \: \ominus \: q {\cal L} }
\: = \: \left( \Lambda_{-q} {\cal L} \right)^{-1}
\end{displaymath}
so we see that 
\begin{displaymath}
\left( S_q {\cal E} \right)^{-1} \: = \: \Lambda_{-q} {\cal E}
\end{displaymath}
for any vector bundle ${\cal E}$, and similarly
\begin{displaymath}
\left( \Lambda_q {\cal E} \right)^{-1} \: = \:
S_{-q} {\cal E}
\end{displaymath}

\section{Thom class formulas}   \label{app:thom}

\subsection{Umkehr maps and the Riemann-Roch formula}

We briefly recall the yoga of Umkehr maps in $K$-theory and ordinary
cohomology, and their relationship through the Riemann-Roch formula.
This story arose in the work of Atiyah, Hirzebruch, and Singer 
as part of the development of index theory, and makes essential use of 
a construction of Pontrjagin and Thom.

\subsection{Thom space}

Let $V\to X$ be a vector bundle.  The \emph{Thom space} of $X$ is the
space 
\begin{equation} \label{eq:10}
X^{V} = D (V)/S (V),
\end{equation}
the disk bundle modulo the sphere bundle.  One needs a metric on $V$
to make sense of $D (V)$ and $S (V)$, but any two metrics give the
same homotopy type for $X^{V}$. 
If $X$ is compact, then we can take $X^{V}$ to be the one-point
compactification of $V.$  Here are some important points about this
construction.

First, if $\uln{n}$ denotes the trivial bundle of rank $n$ over $X,$
then 
\[
     X^{\uln{n}} = \frac{D^{n}\times X}{S^{n-1}\times X} \iso 
\frac{S^{n}\times X}{\ptspace \times X}.
\]
If $Y$ is a pointed space, let $\Sigma^{n}$ denote the $n$-fold
``reduced suspension'' of $Y$: 
\[
   \Sigma^{n} Y = \frac{S^{n}\times Y}{\ptspace \times Y
   \cup S^{n}\times \ptspace}.
\]
If $X_{+}$ refers to the space $X$ with a disjoint point
added, then
\[
   \Sigma^{n} (X_{+}) \iso  \frac{S^{n} \times X}{\ptspace \times X}
   \iso X^{\uln{n}}.
\]
Thus the construction $V/X \mapsto X^{V}$ is a generalization of
reduced suspension.

Second, formation of the Thom spectrum is natural with respect to
pull-backs:  given  
\[
\begin{CD}
@. V \\
@. @VVV \\
Y @> f >> X,
\end{CD}
\]
we have an induced map of Thom spaces 
\begin{equation} \label{eq:16}
   f: Y^{f^{*}V} \to X^{V}.
\end{equation}

Third, the Thom space construction is 
exponential in the sense that for $V\to X$ and $W\to Y$, we have 
\begin{equation}\label{eq:15}
     (X\times Y)^{V\hat{\oplus} W} \heq X^{V} \Smash Y^{W}.
\end{equation}
Here $\Smash$ denotes the ``smash product'': if $A$ and $B$ are two
pointed spaces, then  
\[
    A \Smash B = \frac{A \times B}{A \times \ptspace \cup \ptspace
    \times B}.
\]
Since the internal Whitney sum $V\oplus W$ is the pull-back along the
diagonal 
\[
\begin{CD}
V \oplus W     @>>> V \hat{\oplus} W \\
@VVV @VVV\\
X @> \Delta >> X\times X,
\end{CD}
\]
by combining \eqref{eq:16} and \eqref{eq:15} we conclude that if $V$
and $W$ are vector bundles over $X$, then we have a map of Thom
spectra 
\[
     X^{V\oplus W} \to X^{V}\Smash X^{W}.
\]
From now on we will not distinguish in notation between the internal and
external Whitney sums.  

Fourth, there's no projection map $X^{V}\to X$, because all of
$S (V)$ was crushed to a single point.  However, the ``relative'' diagonal map 
\[
     V \to V\times V \to X \times V
\]
does induce a map 
\begin{equation}\label{eq:25}
     X^{V} \xra{} X_{+} \Smash X^{V}.
\end{equation}
Note that the zero section $\zeta: X\to V$ does give rise to a map
$\zeta: X \to X^{V},$ also called the zero section.

Let $E$ be a generalized cohomology theory.  If $Y$ is a pointed
space, then we can use the inclusion $\ptspace \to Y$ to form the
associated reduced cohomology  
\[
\redE^{*} (Y)  = \Ker E^{*} (Y) \to E^{*} (\ptspace).
\]
The Mayer-Vietoris sequence implies that we have the so-called ``suspension isomorphism'' 
\[
   \redE^{*} (Y)\iso \redE^{*+n} (\Sigma^{n} Y).
\]
If $X$ an unpointed space then 
\[
    E^{*} (X) \iso \redE^{*} (X_{+}),
\]
so the suspension isomorphism can be read to say that 
\begin{equation} \label{eq:17}
    E^{*} (X) \iso \redE^{*} (X_{+}) \iso \redE^{*+n} (X^{\uln{n}}).
\end{equation}

An \emph{orientation} or \emph{Thom isomorphism} for $V$ in $E$-theory
is an isomorphism 
\begin{equation} \label{eq:18}
    E^{*} (X) \iso \redE^{*+r} (X^{V}),
\end{equation}
where $r$ is the rank of $V.$  Comparing \eqref{eq:17}, we see that an
orientation of $V$ is a generalization of the suspension isomorphism.

Typically this terminology arises when $E$ is a ring spectrum:
that is, $E^{*}(X)$ is a graded ring, rather than merely a graded
abelian group.  In that case, the relative diagonal \eqref{eq:25}
gives us a map  
\[
      E^{*} (X)\otimes \redE^{*} (X^{V}) \rightarrow \redE^{*} (X^{V}),
\]
so $\redE^{*} (X^{V})$ is a module over the ring $E^{*} (X).$  We 
then require the Thom isomorphism \eqref{eq:18} to be an
isomorphism of $E^{*} (X)$-modules.  Since $E^{*} (X)$ is free of rank
$1$ as a module over itself, the Thom isomorphism is completely
determined by a choice of generator $U\in \redE^{r} (X^{V}),$ called
the \emph{Thom class}.   

The pull-back of the Thom class $U$ along the zero section $\zeta:
X\to X^{V}$ is called the ``Euler class'' 
\[
     e (V) = \zeta^{*}U \in E^{r} (X).  
\]
The composition 
\[
     E^{*} (X) \xra{\text{Thom}} E^{*+r} (X^{V}) \xra{\zeta^{*}}
     E^{*+r} (X) 
\]
is multiplication by the Euler class.

The name ``orientation'' arises from the fact that in the case that
$E$ is ordinary cohomology with integer coefficients, a choice of Thom
class  is equivalent to a choice of orientation for the vector bundle
$V$, in the classical sense.  If $M$ is a compact oriented manifold
with tangent bundle $TM$, then the Euler class $e (TM)$ has the 
property that 
\[
        \chi (M) = \int_{M} e (TM),
\]
where $\chi (M)$ is the Euler characteristic 
\[
   \chi (M) = \sum_{i} (-1)^{i} \dim H^{i} (M;\Q).
\]

When $E$ is a ring spectrum, one often asks not just for one 
orientation of the single bundle $V/X$, but a compatible family of
orientations of a family of bundles.  

Let $\mathcal{V}$ be a family of vector bundles which contains the
trivial bundles and is closed under pull-back and Whitney sum.  For
example, $\mathcal{V}$ could be the family of complex vector bundles,
or spin vector bundles, etc.     A \emph{(stable exponential)
orientation} of $\mathcal{V}$ in $E$-theory is a rule  $\Phi$ which 
assigns to $V/X \in \mathcal{V}$ of rank $r$ a Thom class 
\[
  \Phi (V/X) \in E^{r} (X^{V}),
\]
which is 
\begin{enumerate}
\item natural, in the sense that 
\[
     f^{*} \Phi (V/X) = \phi (f^{*}V/Y) \in E^{r} (Y^{f^{*}V}),
\]
if $V/X$ and $f: Y\to X$;
\item exponential, in the sense that 
\[
     \Phi (V\oplus W / (X\times Y)) = \phi (V/X) \phi (W/Y) \in E^{r+s}
     ((X\times Y)^{V\oplus W}),
\]
using the equivalence \eqref{eq:15}; and 
\item unital, meaning that the induced Thom isomorphism 
\[
    E^{*} (X) \iso \redE^{*+n} (X^{\uln{n}})
\]
coincides with the suspension isomorphism \eqref{eq:17}.
\end{enumerate}

\subsection{The Pontrjagin-Thom construction}

Orientations in a generalized cohomology theory are an important
ingredient in the theory of generalized integration/intersections.   The
other important ingredient is the Pontrjagin-Thom construction.

First suppose that 
\[
   j: X \to Y
\]
is an embedding, with normal bundle $\nu.$  By the Tubular Neighborhood
Theorem, there's a neighborhood $N$ of $X$ in $Y$ and a homeomorphism 
\[
    h: D (\nu) \iso T
\]
making the diagram 
\[
\xymatrix{
{D (\nu)}
 \ar[rr]^{h}
& &
{T}\\
&
{X}
 \ar[ul]^{\zeta}
 \ar[ur]_{j}
}
\]
commute ($\zeta$ is the zero section of $\nu$).

By collapsing the complement of $T$ to a point, we get a map of
pointed spaces
\begin{equation}\label{eq:21}
   \tau (j):  Y_{+} \to X^{\nu}
\end{equation}
from the one-point compactification of $Y$ to the Thom space of $\nu.$
In our applications, $Y$ will be compact, and in that case $Y_{+}$
indicates $Y$, with a disjoint basepoint added.  This is called the
``Pontrjagin-Thom collapse.''

Now suppose that $f: X\to Y$ is a proper map (for example, a fiber
bundle with compact fiber).  Let's first ``convert it to an
embedding,'' for example by choosing an embedding $X \to \R^{N}$ and
then considering 
\[
       \tilde{f}: X\to Y\times \R^{N}.
\]
More generally you could find some vector bundle $V/Y$ and an
embedding 
\[
          \tilde{f}: X\to V
\]
making the diagram 
\[
\xymatrix{
{X}
\ar[r]^{\tilde{f}}
 \ar[dr]_{f} 
&
{V}
\ar[d]
\\
&
{Y}
}
\]
commute.  Now apply the Pontrjagin-Thom construction: we get a map 
\[
   Y^{\uln{N}} \to X^{\nu (\tilde{f})}
\]
or more generally 
\begin{equation} \label{eq:19}
   Y^{V} \to X^{\nu (\tilde{f})}.
\end{equation}
All this appears to depend on the choice of $\tilde{f}$, but not as
much as you might think.  Note that
\[
     V + TY = TX + \nu (\tilde{f}),
\]
and experience with $K$-theory makes you willing to rewrite this as 
\[
     \nu (\tilde{f}) - V = TY - TX.
\]
If $X\to Y$ is a fiber bundle then, with $Tf$ the bundle of tangents
along the fiber, we have
\[
     TX = TY + Tf
\]
and so 
\[
     TY - TX = - Tf
\]
so in the end we can write 
\[
    \nu (\tilde{f}) - V = - Tf.
\]
A stable homotopy theorist will then tell you that the map
\eqref{eq:19} gives rise to stable map 
\begin{equation}\label{eq:20}
    Y_{+} = Y^{\uln{0}}  \xra{\tau (f)} X^{-Tf}
\end{equation}
whose homotopy class depends only on $f: X\to Y.$

This is a generalization of the Pontrjagin-Thom map of an
embedding: if $j: X \to Y$ is an embedding then 
\[
    Tj = - \nu (j)
\]
and so the map in \eqref{eq:20} can equivalently be written 
\[
   \tau (j): Y_{+} \to X^{\nu (j)},
\]
and as such is the same as the map in \eqref{eq:21}.

\subsection{The Umkehr map}

Suppose that 
\[
    f: X\to Y
\]
is a proper map and $E$ is a ring spectrum.   Let $d=\dim X - \dim Y$
to $\dim Tf = d.$  The Pontrjagin-Thom
construction gives a map 
\[
     \tau (f): Y_{+} \to X^{-Tf},
\]
and so in $E$-theory we get a homomorphism 
\begin{equation}\label{eq:22}
   \tau (f)^{*}:    \redE^{*} (X^{-Tf}) \to \redE^{*} (Y_{+}) \iso E^{*} (Y).
\end{equation}

If $-Tf$ or equivalently $Tf$ is \emph{oriented} in $E$-theory, then
we have a Thom isomorphism 
\[
   \Phi: E^{*} (X) \iso \redE^{*-d} (X^{-Tf})
\]
and composing with the Pontrjagin-Thom map we get finally a
homomorphism 
\begin{equation} \label{eq:23}
     f_{\Phi} =  f_{!}: E^{*} (X) \to E^{*-d} (Y).
\end{equation}
This is the ``generalized integration map'' in $E$-theory.  When $E$
is ordinary cohomology, then via the De Rham isomorphism $f_{!}$
corresponds to integration of differential forms over the fiber.  
The Atiyah-Singer index theorem interprets $K$-theory's $f_{!}$ as
the index of a families elliptic  operator.  

The notation $f_{!}$ or $f_{*}$ 
is fairly standard: see for example
\cite{AB:MomentMap,Dyer:Cohomology}.  One problem with the notation is
that it does not indicate the dependence on the orientation $\Phi$:
this is like writing $\int f$ without indicating the volume form.
In this paper we'll instead write $f_{\Phi}$, to indicate the choice
of Thom isomorphism.

In our applications, we will suppose that $\Phi$ is a stable, natural,
and exponential family of orientations for complex vector
bundles in $K$-theory.   The compatibility of the family of
orientations $\Phi$ implies that the associated Umkehr maps enjoy the
following properties.

\begin{enumerate}
\item Suppose $f: X \to Y$ and $g: Y\to Z$ are complex-oriented maps.
We have an exact sequence 
\[
        0 \rightarrow Tf \rightarrow Tgf \rightarrow f^{*} Tg
        \rightarrow 0.
\]
If we allow $Tgf$ to inherit a complex structure from $Tf$ and $Tg$, 
then 
\begin{equation} \label{eq:4}
            g_{\Phi} f_{\Phi} = (gf)_{\Phi}.
\end{equation}
\item If 
\[
\begin{CD}
W @> i >> X \\
@V g VV @VV f V \\
Y @> j >> Z
\end{CD}
\]
is a pull-back diagram, then 
\begin{equation} \label{eq:2}
       g_{\Phi} i^{*} = j^{*} f_{\Phi}: K (X) \to K (Y;R).
\end{equation}
\item If $V\to X$ is a complex vector bundle, and 
\[
   \zeta: X  \to V
\]
is its zero section, then 
\[
   \zeta_{\Phi}: E^{*} (X) \to \redE^{*+r} (X^{V})
\]
is the Thom isomorphism, and so 
\begin{equation} \label{eq:3}
   \zeta^{*} \zeta_{\Phi} (u) = u \cdot e_{\Phi} (V),
\end{equation}
where $e_{\Phi} (V) = \zeta^{*}\zeta_{\Phi} (1)$ is the Euler class
associated to the vector bundle $V$ and orientation $\Phi.$
\end{enumerate}

If $X$ is a manifold of dimension $d$, let $\pi^{X}$ be the map to a
point $\pi^{X}: X\to \ast.$  A complex structure on $\pi^{X}$ is a complex structure
on $TX,$ and given such a complex structure we can form 
\[
   \pi^{X}_{\Phi} : E^{*} (X) \to E^{*-d} (\ptspace),
\]
and so 
\[
    \pi^{X}_{\Phi} (1) \in E^{-d} (\ptspace).
\]
By Thom's theory of bordism, the rule 
\[
      X \mapsto \pi^{X}_{\Phi} (1)
\]
is a ring homomorphism 
\[
    \Omega^{U}_{*} \to E^{-*} (\ptspace)
\]
from the bordism ring of stably complex manifolds to $E^{*} (\ptspace).$  It is
called the \emph{genus} associated to the orientation  $\Phi.$

Let $V\to X$ be a complex vector bundle over $X.$  Let $s$ be a section
of $V$, which intersects the zero section $\zeta$ transversely.  Let
$Z = s^{-1} (0)$, so we have a pull-back diagram 
\[
\begin{CD}
Z @> j >> Y \\
@Vj VV @VV \zeta V   \\
Y @> s >> V.
\end{CD}
\]
Using the rules above we have 
\begin{equation} 
       j_{\Phi}j^{*}u = s^{*}\zeta_{\Phi}u.
\end{equation}
But $s^{*} = \zeta^{*}$ since $s$ and $\zeta$ are homotopic, and so 
we have 

\begin{equation} \label{eq:5}
      j_{\Phi}j^{*} u = \zeta^{*}\zeta_{!}u = u  \cdot e_{\Phi} (V).
\end{equation}

The ``topological Riemann-Roch formula'' studies how the Umkehr
changes as the family of orientations $\Phi$ changes.  Usually this is
expressed in terms of a change of cohomology theories.  We shall only
need the case of ordinary cohomology and $K$-theory, so we state it in
that case.   Let 
\[
     \ch: K (X) \to H^{\text{even}} (X;\Q)
\]
be the Chern character.

Let $\Phi$ be a multiplicative family of orientations in $K$-theory.
By the splitting principle, the following rules determine an
exponential characteristic class for complex 
vector bundles 
\[
     V/X \mapsto \mathcal{C} (V/X) \in H^{\text{even}} (X;\Q):
\]
\begin{enumerate}
\item $\mathcal{C} (\uln{n}) = 1$;
\item $\mathcal{C} (V\oplus W) = \mathcal{C} (V) \mathcal{C} (W)$ for
any complex $V,W$ over $X$;
\item If $L$ is a complex line bundle, then 
\[
\mathcal{C} (L) = \frac{c_{1}L}{\ch e_{\Phi} (L)}.
\]
\end{enumerate}

Then we have the following, which you can glean from \cite{MR38:5245}
and which is stated in explicit form in \cite{Dyer:Cohomology}.

\begin{Proposition}\label{t-pr-t-r-r}
Let $f: X \to Y$  be a proper complex oriented, and let $\Phi$ be a
stable exponential family of complex orientations in $K$-theory.  Then 
\[
     \ch f_{\Phi} (u) = \int_{f} \mathcal{C} (Tf) \ch u.
\] %\qed
\hfill $\Box$   % Try this as substitute for qed marker.
\end{Proposition}

\begin{Example}
There is a family $\lambda$ of complex orientations for $K$-theory with the
property that 
\[
  e_{\lambda} (V) = \Lambda_{-1} (\overline{V}).
\]
The associated characteristic class is the Todd class
\[
   \Td  (V) = \prod_{i} \frac{x_{i}}{1 - e^{-x_{i}}},
\]
where the $x_{i}$ are defined by 
\[
   c (V) = \prod_{i} (1+x_{i}).
\]
The genus 
\[
     \pi_{\lambda}^{X} (1) = \int_{X}\Td (TX)
\]
is the Todd genus.
\end{Example}

\begin{Example}\label{ex-1}
Now suppose that $R$ is a ring, and 
\[
\mu: K (X) \to K (X;R)^{\times}
\]
is a characteristic class of complex vector
bundles satisfying 
\[
     \mu (\uln{n}) = 1 
\]
and 
\[
     \mu (V+W) = \mu (V) \mu (W).
\]
Then $\mu$ determines a stable exponential family $\Phi$ of orientations  for
complex vector bundles in $K$-theory by the formula 
\[
      \Phi (V) = \lambda (V)\mu (V).
\]
Now the Riemann-Roch formula gives 
\begin{equation}\label{eq:24}
      \ch f_{\Phi} (u) = \int_{f} \Td (Tf) \ch (\mu (V)^{-1} u).
\end{equation}
\end{Example}

\begin{Example}\label{ex-2}
For example, we can set $R = \Z\psb{q}$ and
\[
   \mu (V) = \prod_{n\geq 1}\Lambda_{-q^{n}} (V^{\C}).
\]
We write $\sigma$ for the family of orientations of complex vector
bundles given by 
\[
     \sigma (V) = \lambda (V)\bigotimes_{n=1,2,3,\dotsc} \Lambda_{-q^{n}} (V^{\C}).
\]
Suppose that 
\[
   c (TX) = \prod_{i} (1+x_{i}).  
\]
The resulting genus is 
\begin{align*}
   \pi_{\Phi}^{X} (1) & =  
\int_{X}\Td (X) 
\ch\left(\bigotimes_{n=1,2,3,\dotsc} S_{q^{n}} ((TX)^{\C})\right)\\
& = 
\int_{X}\Td (X) \prod_{i}\prod_{n\geq
   1}\frac{1}{(1-q^{n}e^{x_{i}}) (1-q^{n}e^{-x_{i}})}.
\end{align*}
If $X$ is an $SU$-manifold so that its A-hat class and Todd class
coincide, then (up to factor depending only on the dimension of $X$)
this is the ``Witten genus'' of $X.$ 
%\cite{} \textbf{Give modularity properties}
\end{Example}

\subsection{Thom classes for Landau-Ginzburg models}
\label{app:thom:lg}

Let $z$ denote the standard complex representation of $S^{1}$, and so
the complex representation ring of $S^{1}$ is $R[S^{1}]\iso
\Z[z,z^{-1}].$  Let $A$ denote $S^{1}$-equivariant $K$-theory with
coefficients in the ring $\Z\psb{q^{1/2}},$ so 
\[
     A (\ptspace) \iso R[S^{1}]\psb{q^{1/2}} \iso \Z[z,z^{-1}]\psb{q^{1/2}}.
\]
If $V$ is a complex vector bundle, let $z^{n}V$ be $V$ considered as
an $S^{1}$-equivariant vector bundle with the circle acting by $z\mapsto z^{n}.$

Let 
\[
\mu: K (X) \to A (X)^{\times}
\]
be the exponential characteristic class given by 
\[
 \mu (V) =  \bigotimes_{k=1/2,3/2,\dots} S_{-q^{n}} ((z^{-1}V)^{\C}).
\]
$\Phi$ be the orientation for complex vector bundles in $A$-theory
given by 
\[
   \Phi (V) = \sigma (V) \mu (V),
\]
where $\sigma$ is the orientation of Example \ref{ex-2}.  If $Y$ is a
manifold with a complex tangent 
bundle, then the Riemann-Roch formula gives
\begin{equation}\label{eq:7}
     \pi^{Y}_{\Phi} (x) = \int_{Y} \Td (TY) \ch 
\left(x \bigotimes_{n=1,2,3,\dots}S_{q^{n}} (TY^{\C})
 \bigotimes_{n=1/2,3/2,\dotsc}\Lambda_{q^{n}} ((z^{-1}TY)^{\C})\right).
\end{equation}
Setting $x=1$ gives the elliptic genus on $Y$ 
discussed in section~\ref{22quintic}; see \eqref{eq:14}.  

Now suppose that $\cG$ is a complex vector bundle over a manifold
$B$.  Let $\zeta$ denote its zero section, and suppose that $G$ is
another section which intersects $\zeta$ transversely.  Let $Y = G^{-1}
(0)$ so that we have a pull-back diagram of the form 
\begin{equation} \label{eq:6}
\begin{CD}
Y @>j >> B \\
@V j VV @VV \zeta V \\
B @> s >> \cG,
\end{CD}
\end{equation}
In this situation 
\[
     \nu (j) = j^{*}\nu (\zeta) = \cG\restr{Y}.  
\]
We have a commutative diagram 
\[
\xymatrix{
{Y}
 \ar[r]^-{j}
 \ar[dr]_-{\pi^{Y}}
&
{B} 
\ar[d]^{\pi^{B}}
\\
&
{\ast},}
\]
and so equation \eqref{eq:4} implies that 
\[
   \pi^{Y}_{\Phi} (x) = \pi^{B}_{\Phi} j_{\Phi} (x).
\]
We shall apply this formula in the case that 
\[
           x = j^{*}u,
\]
in which case \eqref{eq:5} implies that 
\[
    j_{\Phi} j^{*}u = u e_{\Phi} (\cG) =  ue_{\sigma} (\cG)\mu (\cG)
    =  u\Lambda_{-1} (\overline{\cG})\bigotimes_{n=1,2,\dotsc}
    \Lambda_{-q^{n}} (\cG^{\C})
    \bigotimes_{n=1/2,3/2,\dotsc} S_{-q^{n}} ((z^{-1} \cG)^{\C})
\]
so that 
\begin{align}
    \pi^{Y}_{\Phi}  (j^{*}u)  = &
    \pi^{B}_{\Phi}j_{\Phi} (j^{*}u) \notag \\
   = &
    \pi^{B}_{\Phi}\left( u\Lambda_{-1} (\overline{\cG}) \Phi (\cG) \right) \notag \\
 = &
    \int_{B} \Td (TB) \wedge
\ch \left(u\bigotimes_{n=1,2,3,\dots}S_{q^{n}} (TB^{\C})
\bigotimes_{k=1/2,3/2,\dots} \Lambda_{q^{n}} ((z^{-1} TB)^{\C}) \right) \notag
\\
& \wedge 
\ch\left( \Lambda_{-1} (\overline{\cG}) \bigotimes_{n=1,2,3,\dots}
\Lambda_{-q^{n}} (\cG^{\C})
\bigotimes_{n=1/2,3/2,\dots} S_{-q^{n}} ((z^{-1}\cG)^{\C})\right). \label{eq:8}
\end{align}
Recalling that for a complex vector bundle $V$ we have $V^{\C}\iso V
\oplus \Bar{V}$, we see that this expression matches
\eqref{genus-genlci}.

\begin{Example}
The situation in \S\ref{22quintic} arises when $Y$ is a generic
quintic hypersurface in $P=\P^{4}$: so $s$ is a generic quintic and
$\cG = \O (5)$.  Then 
\begin{align*}
\overline{\cG} & = \O (-5) \\
\intertext{and}
\lambda (\cG) & = \Lambda_{-1} (\overline{\cG}) = 1 - \O (-5).
\end{align*}
Also setting $u=1$ and comparing with \eqref{eq:7}, \eqref{eq:8} becomes 
\begin{multline}
\int_{Y} \Td (TY) \ch
\left(\bigotimes_{n=1,2,3,\dots}S_{q^{n}} (TY^{\C})
\bigotimes_{k=1/2,3/2,\dots} \Lambda_{q^{n}} ((zTY)^{\C})\right) = \\
  \int_{P} \Td (TP) \ch \left(\bigotimes_{n=1,2,3,\dots}S_{q^{n}} (TP^{\C})
\bigotimes_{k=1/2,3/2,\dots} \Lambda_{q^{n}} ((z TP)^{\C}) \right) 
\\
  \ch\left( (1-\O (-5)) \bigotimes_{n=1,2,3,\dots}\Lambda_{-q^{n}} (\O (5)^{\C})
 \bigotimes_{k=1/2,3/2,\dots} S_{-q^{n}} (z(\O (5))^{\C})\right). \label{eq:26}
\end{multline}
One checks easily that \eqref{eq:26} implies 
that the genera \eqref{eq:7} and \eqref{eq:8} coincide.
\end{Example}

\subsection{The formula for a complex of vector bundles}
\label{app:thom:vb}

We continue to suppose that $\cG$ is a line bundle on $B$, with a
section $G$ which intersects $\zeta$ transversely.   We let $Y=G^{-1}
(0)$, and we write $j: Y\to B$ for the inclusion.  Once again we have 
$\nu (j) = \cG.$

Now we suppose 
that
\[
     \cF = \left(\cF_{0} \xrightarrow{s_{0}} \cF_{1} \xrightarrow{s_{1}} \cF_{2}\right)
\]
is a complex of vector bundles on $B:$ we suppose that $s_{i}s_{i-1} =
0$, but we do not suppose that the complex is exact.    Let 
\begin{align*}
    \cE_{0}& = \Ker s_{0} \\
    \cE_{1} &= \Ker s_{1}/\Img s_{0} \\
    \cE_{2} & = \cF_{2}/ \Img s_{1}.
\end{align*}
In the physical situation studied in
\S\ref{sec:models-coh-of-monads} it is convenient to reverse signs and
set 
\[
     \cE' = - \cE_{0} + \cE_{1}- \cE_{2}
\]

If $V_{0},\dotsc ,V_{n}$ is a sequence of complex vector bundles, then
we define 
\[
   \Lambda_{t} (V) = \Lambda_{t} (\sum_{i} (-1)^{i}V_{i})\iso
   \bigotimes \Lambda_{t} (V_{i})^{(-1)^{i}} 
\]
and similarly for $S_{t}V$.

\begin{Lemma}
We have 
\[
    \cF_{0} - \cF_{1} + \cF_{2} = \cE_{0} - \cE_{1} + \cE_{2}
\]
in $K (B)$, and so 
\begin{equation} \label{eq:1}
      S_{-q^{n}} (z^{-1} \cF)  = S_{-q^{n}} (z^{-1} \cE) = \Lambda_{q^{n}} (z^{-1}\cE')
\end{equation}
in $K_{S^{1}} (B;\Z\psb{q}).$
\end{Lemma}

%\begin{proof}
{\it Proof.}
The sections $s_{i}$ lead to  decompositions
\begin{align*}
\cF_{0} &\iso \Ker s_{0} + (\Ker s_{0})^{\perp} \iso \cE_{0} + \Img s_{0}\\
\cF_{1} &\iso  \Ker s_{1} + (\Ker s_{1})^{\perp} \iso 
\Img s_{0} + \cE_{1} + \Img s_{1} \\
     \cF_{2} &\iso \Img s_{1} + \cE_{2}.
\end{align*}
\hfill  $\Box$
%\end{proof}

\begin{Proposition}
In this situation, we have  
\begin{multline*}
 \int_{Y}\Td (Y)\wedge 
\ch \left( \bigotimes_{n=1,2,\dotsb} S_{q^{n}} \left((TY)^{\C}\right)
           \bigotimes_{n=1/2,3/2,\dotsb} \Lambda_{q^{n}} ((z^{-1}\cE')^{\C})
    \right)  \\
= \: \int_{B}\Td (B) \wedge
\ch \left( \bigotimes_{n=1,2,\dotsb} S_{q^{n}} \left((TB)^{\C}\right)
      \bigotimes \Lambda_{-1} (\Bar{\cG}) \bigotimes_{n=1,2,\dotsc} \Lambda_{-q^
{n}} (\cG^{\C})
\right. \\
\hspace*{1.5in}  \left.
           \bigotimes_{n=1/2,3/2,\dotsb} S_{-q^{n}}
	   ((z^{-1}\cF)^{\C})\right). 
\end{multline*}
\end{Proposition} 

In particular this shows the elliptic genera \eqref{eg:nlsm:monad:ci}
and \eqref{genus-genlcimonad} coincide.

%\begin{proof}
{\it Proof.}
We calculate the pushforward using the multiplicative 
orientation  $\sigma$ of Example \ref{ex-2},  whose Euler class is 
\[
    e_{\sigma} (V) = \Lambda_{-1} (\Bar{V}) \bigotimes_{n\geq 1}
    \Lambda_{-q^{n}} (V^{\C}).  
\]
As always we have
\begin{equation} \label{eq:11}
    \pi^{Y}_{\sigma} ( x) = \pi^{B}_{\sigma} (j_{\sigma} (x)).
\end{equation}
We apply this formula to 
\[
x=j^{*}u = j^{*}\left(\bigotimes_{n=1/2,3/2,\dotsc }
S_{-q^{n}}(z^{-1}\cE)^{\C}\right), 
\]
so
\begin{equation} \label{eq:12}
    j_{\sigma} j^{*} u = u e_{\sigma} (\cG).
\end{equation}
using \eqref{eq:5}.    Thus 
\begin{equation} \label{eq:13}
    \pi^{Y}_{\sigma} 
\left( \bigotimes_{n=1/2,3/2,\dotsc }
S_{-q^{n}}(z^{-1}\cE)^{\C}\right) = 
\pi_{\sigma}^{B}\left( \Lambda_{-1} (\Bar{\cG})\bigotimes_{n\geq 1}
    \Lambda_{-q^{n}} (\cG^{\C}) 
 \bigotimes_{n=1/2,3/2,\dotsc }S_{-q^{n}}(z\cE)^{\C}
          \right).
\end{equation}
Using the Lemma, we may replace $S_{-q^{n}} ((z^{-1}\cE)^{\C})$ with
$S_{-q^{n}} ((z^{-1}\cF)^{\C})$ or $\Lambda_{q^{n}}
((z^{-1}\cE')^{\C})$.  Doing so and  
using the Riemann-Roch formula (Proposition \ref{t-pr-t-r-r}) to 
rewrite both sides as integrals after applying the Chern character gives
the equation in the statement of the Proposition.

\hfill $\Box$

\subsection{R-sector genera}\label{sec:r-sector-genera}

It is easy to adapt the methods of the preceding sections to the
R-sector genera studied in this paper.  For definiteness we compare the
genera \eqref{eq:41} and \eqref{eq:37}, and we describe the essential
idea in a slightly different way from the preceding sections.  Let $Y$
be a submanifold of $B$ with normal bundle $\mathcal{G}$.  The
intersection theory argument we have been using shows in general that
\[
   \int_{Y}\Td (Y)\ch \left(\xi (TY)\right) = 
   \int_{B}\Td (B) \ch\left(\Lambda_{-1} (\mathcal{G}^{\vee})\xi
   (\mathcal{G})^{-1}\xi (TB)\right),
\]
and 
\[
    \int_{Y}\hat{A} (Y)\ch \left(\xi (TY)\right) = 
   \int_{B}\hat{A}(B) \ch\left((\Delta_{+} (\mathcal{G}^{\vee}) - \Delta_{-} (\mathcal{G}^{\vee}))\xi
   (\mathcal{G}^{\vee})^{-1}\xi (TB)\right).
\]
Here $\xi$ is an exponential function
\[
     KO (X) \to K (X;R)^{\times}
\]
for some ring of coefficients $R$.  

For example, consider the R-sector NLSM genus \eqref{eq:41}
\begin{displaymath}
\int_Y \Td(TY) \wedge \ch\left( 
z^{-y/2} \left( \det TY \right)^{1/2} \Lambda_1( z T^* Y) 
\bigotimes_{n=1,2,3,\cdots} S_{q^n}( (TY)^{\bf C} )
\bigotimes_{n=1,2,3,\cdots} \Lambda_{q^n}( (z^{-1} TY)^{\bf C} )
\right)
\end{displaymath}
and its Landau-Ginzburg couterpart \eqref{eq:37}
\begin{eqnarray}
\lefteqn{
q^{0}
} \nonumber \\
& & \cdot \int_B \Td(TB) \wedge \ch\Biggl(
z^{+b/2} z^{-(b-y)/2} \Lambda_1( z^{-1} TB) \otimes 
\Lambda_{-1}( {\cal G}^{\vee} ) \otimes
%\Lambda_{-1}( TB ) \otimes
\Lambda_1( z {\cal G}^{\vee} )
\nonumber \\
& & \hspace*{1.5in} \cdot
\left( \det T^*B \right)^{1/2}
\left( \det {\cal G}^{\vee} \right)^{-1/2}
\nonumber \\
& & \hspace*{1.5in} \cdot
\bigotimes_{n=1,2,3,\cdots} S_{q^n}( (TB)^{\bf C} )
\bigotimes_{n=0,1,2,\cdots} S_{-q^n}( (z {\cal G}^{\vee} )^{\bf C} )
\nonumber \\
& & \hspace*{1.5in} \left. \cdot
\bigotimes_{n=1,2,3,\cdots} \Lambda_{q^n}( (z^{-1} TB)^{\bf C} )
\bigotimes_{n=1,2,3,\cdots} \Lambda_{-q^n}( ({\cal G}^{\vee})^{\bf C} )
\right).
\end{eqnarray}
The factor  
\[
z^{-y/2}\det (TY)^{1/2}\Lambda_{1} (zT^{*}Y)\iso 
z^{y/2}\det (T^{*}Y)^{1/2}\Lambda_{1} (z^{-1}TY)
\]
in the $Y$ integral corresponds to a
factor 
\[
z^{b/2}\det (T^{*}B)^{1/2}\Lambda_{1} (z^{-1}TB)z^{- (b-y)/2}\det (\mathcal{G}^{\vee})^{-1/2}S_{-1} (z^{-1}\mathcal{G}),
\]
in the $B$
integral.  
The $S_{-1}$ part comes from the $n=0$ case of $S_{-q^{n}}
((z\mathcal{G}^{\vee})^{\C}),$ which contributes
\[
    S_{-1} (z\mathcal{G}^{\vee})\otimes S_{-1} (z^{-1}\mathcal{G}).
\]
The second factor is the one we need, while the first is cancelled by
the factor $\Lambda_{-1} (z\mathcal{G}^{\vee})$ in the first line of
the $B$ integral.

Similar remarks apply to the R-sector genera involving complexes of
vector bundles; the proof is left to the interested reader.

\end{document}